\documentclass[twocolumn,pra,floats,aps,amsmath,amssymb,superscriptaddress,nofootinbib]{revtex4-1} 
\usepackage{amsmath,amssymb,amsthm,mathrsfs,amsfonts,dsfont,amstext} 
\usepackage{cases} 
\usepackage{textcomp,pbox}
\usepackage[export]{adjustbox}
\usepackage{bm}
\usepackage{dcolumn,booktabs,url}
\usepackage[scaled]{helvet}
\usepackage{sansmath,gensymb}
\usepackage{tikz,graphicx,transparent,color}
\usepackage{multirow}
\usepackage[separate-uncertainty = true]{siunitx}
\usepackage{comment}
\usepackage{physics}
\usepackage{pdfpages}
\usepackage{array}
\usepackage{tabularx}
    \usepackage{empheq}
    \usepackage{xcolor}
    \usepackage[most]{tcolorbox}

\usepackage{import}
\usepackage{xifthen}
\usepackage{pdfpages}
\usepackage{transparent}
\usepackage{svg}
\usepackage{adjustbox}

\makeatletter
\AtBeginDocument{\let\LS@rot\@undefined}
\makeatother

\newcolumntype{C}[1]{>{\centering\let\newline\\\arraybackslash\hspace{0pt}}m{#1}}

\usepackage[colorlinks=true]{hyperref}
\usepackage{graphicx}

\hypersetup{
     colorlinks   = true,
     citecolor    = red,
     linkcolor    = blue,
     urlcolor     = red     
}

\graphicspath{{./figs/}}

\begin{document}

\newcommand{\TitleName}{Quantum state transfer between a frequency-encoded photonic qubit and a quantum dot spin in a nanophotonic waveguide}
\title{\TitleName}

\newcommand{\AffCPH}{Center for Hybrid Quantum Networks (Hy-Q), The Niels Bohr Institute, University~of~Copenhagen,  DK-2100  Copenhagen~{\O}, Denmark}

\newcommand{\AffWeiz}{Chemical Physics Department, Weizmann Institute of Science, 76100 Rehovot, Israel}

\author{Ming~Lai~Chan$^{\dagger,}$}
\thanks{These authors contributed equally to this work.}
\affiliation{\AffCPH{}}
\def\thefootnote{$\dagger$}\footnotetext{ming-lai.chan@nbi.ku.dk}

\author{Ziv~Aqua}
\thanks{These authors contributed equally to this work.}
\affiliation{\AffWeiz{}}

\author{Alexey~Tiranov}
\affiliation{\AffCPH{}}
\author{Barak~Dayan}
\affiliation{\AffWeiz{}}
\author{Peter~Lodahl}
\affiliation{\AffCPH{}}
\author{Anders S. Sørensen}

\affiliation{\AffCPH{}}

\date{\today}

\begin{abstract}

We propose a deterministic yet fully passive scheme to transfer the quantum state from a frequency-encoded photon to the spin of a quantum-dot mediated by a nanophotonic waveguide. We assess the quality of the state transfer by studying the effects of all relevant experimental imperfections on the state-transfer fidelity. We show that a transfer fidelity exceeding $95\%$ is achievable for experimentally realistic parameters. Our work sets the stage for deterministic solid-state quantum networks tailored to frequency-encoded photonic qubits. 


\end{abstract}

\maketitle

\section{Introduction}

Optical photons are  regarded as the ``ideal courier" for transmitting and distributing information across a quantum network. For long-distance distribution of quantum information, an efficient light-matter interface is required whereby single photons and single quantum emitters are coherently coupled~\cite{Kimble2008,Northup2014,Lodahl2017b,Wehner2018,Kim2020,Reiserer2015}. Such interfaces can be harnessed to construct a general-purpose ``toolbox of quantum photonic devices", which provides the essential  functionalities for quantum networks \cite{Uppu2021}. Examples of functionalities include the ability to coherently exchange quantum states between the emitter and a photon \cite{Northup2014} at the network nodes, the generation of on-demand indistinguishable single photons \cite{Uppu2020} or high-fidelity entangled states \cite{Schwartz2016,Appel2021,Atature2018,Gao2012} for entanglement distribution, or the realization of deterministic quantum logic between photons \cite{Reiserer2015} for quantum computation.



Here we add to this toolbox by developing a theory for the implementation of a two-qubit SWAP gate between a flying photon and a single quantum dot (QD) embedded in a nanophotonic waveguide. This system, which is today a mature platform~\cite{Lodahl2015}, enables near-unity light-matter coupling efficiency \cite{Arcari2014}, thus making the photon-to-emitter state transfer practically deterministic. The considered photonic qubit is encoded in two distinct frequencies, making it compatible with frequency multiplexing in fiber networks \cite{Armstrong2012,Roslund2014,Chen2014,Grimau2017,Joshi2018}. 
We lay out a complete photon-spin quantum-state transfer protocol and analyze all relevant experimental imperfections. 
We present both a perturbative analytical theory, essential for unravelling the importance of the different mechanisms, as well as a complete numerical analysis applicable also beyond the perturbative approximation.

The underlying principle of our protocol is single-photon Raman interaction (SPRINT)~\cite{Rosenblum2017,Bechler2018}, which exploits passive Raman spin flip of the emitter's initial state due to the destructive interference between the incident and scattered fields in a waveguide geometry.
SPRINT was first considered by Pinotsi and Imamoglu~\cite{pinotsi2008single} as an ideal absorber of a single photon. Subsequent theoretical works~\cite{Lin2009heralded,Witthaut2010,Koshino2010deterministic,Bradford2012single,Koshino2013theory,Rosenblum2017, Adrien2020} established that it acts as a photon–atom swap gate and accordingly serves as a quantum memory, which includes a heralding mechanism that validates the arrival of the input photon. It was experimentally demonstrated with a single atom implementing a single-photon router~\cite{shomroni2014}, extraction of a single photon from an optical pulse~\cite{rosenblum2016}, and a photon–atom qubit swap gate~\cite{Bechler2018}. In superconducting circuits it was demonstrated as well~\cite{Inomata2014}, harnessed for highly efficient detection of single microwave photons~\cite{Inomata2016}, and considered for remote entanglement generation~\cite{Koshino2017}. While SPRINT has been performed on various systems, its implementation on a deterministic QD-waveguide platform with inclusion of all relevant physical processes pertaining to QD has not been previously considered.

\begin{figure}[ht]
    \centering
    \includegraphics[width=\linewidth]{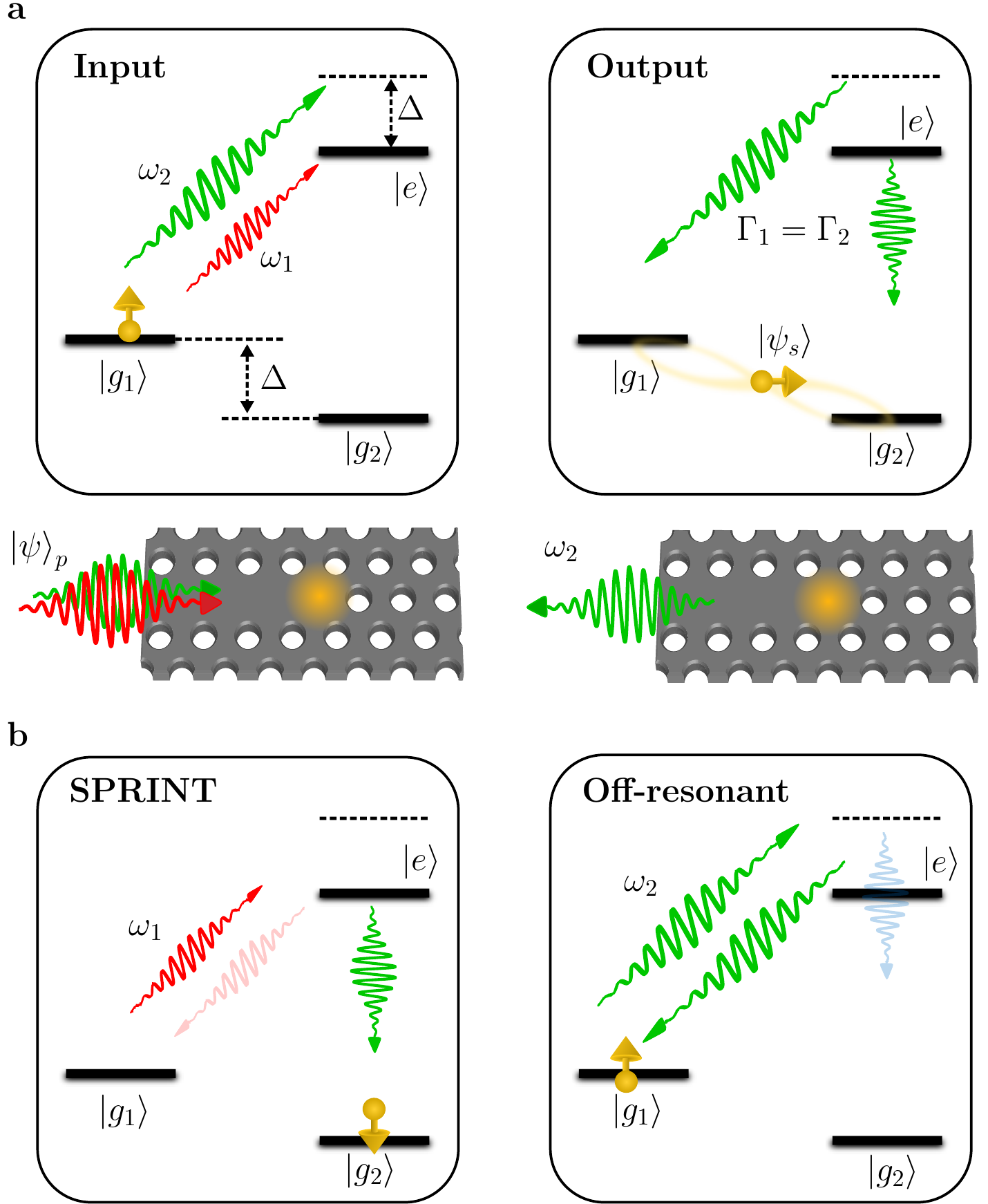}
    \caption{Schematics of the photon-to-spin state transfer in a $\Lambda$-level emitter coupled to a one-sided nanophotonic waveguide. \textbf{(a)} \textbf{Left:} Initial state of the spin-photon system, where the superposition state $\ket{\psi_p}$ is encoded in frequency bins $\ket{\omega_1}$ (red) and $\ket{\omega_2}$ (green) separated by the ground-state splitting $\Delta$. The photonic qubit scatters off an emitter (orange) initialized in state $\ket{g_1}$. \textbf{Right:} the photonic state $\ket{\psi_p}$ is deterministically transferred to the spin carrier $\ket{\psi_s}$. With symmetric decay rates $\Gamma_1=\Gamma_2$ the final spin state corresponds to the incoming photonic state due to the  superposition of SPRINT and off-resonant scattering.   \textbf{(b)} \textbf{Left:} SPRINT, where a resonant photon flips the spin state from $\ket{g_1}$ to $\ket{g_2}$ (solid green). Imperfections induce a small probability to decay to the original spin state (pink). \textbf{Right:} Off-resonant scattering, where the  photon  is almost unaffected (solid green). There is a small probability to flip the spin (light blue) due to finite $\Delta$. }    \label{fig:schematics}
\end{figure}

The operational principle of the quantum-state transfer protocol in a one-sided nanophotonic waveguide is displayed in Fig.~\ref{fig:schematics}(a). The QD features a $\Lambda$-type level structure with  an excited state $\ket{e}$ that decays to two meta-stable spin states $\ket{g_1}$  and $\ket{g_2}$ with  decay rates $\Gamma_1$ and $\Gamma_2$ respectively. State transfer proceeds by exploiting a photonic frequency qubit consisting of a superposition of two frequency states $\ket\omega_1$ and $\ket\omega_2$, where $\omega_1$ and $\omega_2$ denote the frequency of the photon. The frequency $\omega_1$ is resonant with the QD transition between $\ket{g_1}$ and $\ket{e}$ and  $\omega_2$ is resonant with the transition between $\ket{g_2}$ and $\ket{e}$ corresponding to a frequency difference  equal to the ground-state splitting $\omega_2-\omega_1=\Delta$. The QD is initially prepared in state $\ket{g_1}$  and scatters either of the two optical frequencies. 
For optimum operation, the decay rates for the two QD transitions should be equal $(\Gamma_1 = \Gamma_2)$, which is naturally the case for a QD in a bulk sample subject to an in-plane magnetic field (Voigt geometry)~\cite{Warburton:2013aa}. Photonic nanostructures, however, generally introduce a decay rate asymmetry as controlled by the projected local density of optical states~\cite{Appel2020}. Proper spatial positioning of the QD in the waveguide \cite{Pregnolato2020} would therefore be required to meet the symmetry condition, but this is still compatible with the near-unity coupling efficiency~\cite{Lodahl2015}. The dynamics of the state-transfer is most easily understood by considering two separate \textit{cases}, corresponding to each of the two possible incoming frequencies:

\textit{Case 1}: an incoming photon at frequency $\omega_1$ resonantly drives the initial spin state $\ket{g_1}$ to the excited state $\ket{e}$. For $\Gamma_1=\Gamma_2$, the excited state has an equal probability to decay on either of the transitions to the two spin ground states. In the ideal limit of a deterministic and fully coherent photon-emitter interface (i.e. high photonic $\beta$-factor and low decoherence \cite{Uppu2020}, the incident and scattered fields interfere destructively~\cite{Shen2005}. As a consequence only a photon at frequency $\omega_2$ is emitted and the spin state will be deterministically toggled to the state $\ket{g_2}$  \cite{Witthaut2010}, cf. left illustration of Fig.~\ref{fig:schematics}(b).
This is a Raman process driven by a single photon, i.e. the SPRINT interaction.

\textit{Case 2}: the incoming photon of frequency $\omega_2$ is detuned from the QD transition by the ground-state splitting $\Delta$, which leads to only a small probability to excite the QD and therefore to drive the Raman transition. Ideally, i.e., for a large detuning, the incident photon does not interact with the QD and is thus fully reflected back from the one-sided waveguide sample, cf. the right box of Fig.~\ref{fig:schematics}(b).


By combining the two processes, i.e. for an input photon pulse prepared in a superposition of the two carrier frequencies $\ket{\psi_p} = c_a\ket{\omega_1}+c_b\ket{\omega_2}$ (Fig.~\ref{fig:schematics}(a)), a separable frequency-spin state is generated, as  described by the  input-output relation
\begin{align}
    \ket{\psi_p}\otimes\ket{g_1}
    \quad \xrightarrow{\text{ideal}}& \quad\ket{\omega_2} \otimes \ket{\psi_s},
    \label{eq:ideal}
\end{align}
where the output spin qubit $\ket{\psi_s}=i\hat{Y}(c_a \ket{g_1}+c_b \ket{g_2})$ is equivalent to the input quantum state up to a local Pauli-Y rotation. Effectively this operation maps the initial photonic state onto the QD spin. This process is passive as no active control fields are required to trigger the  scattering process. Moreover, by heralding the gate upon the detection of an output photon, our scheme becomes very robust to the considered imperfections. Since the outgoing photon is always supposed to have a frequency $\omega_2$ this robustness can be further enhanced by adding a frequency filter which removes photons of the wrong frequency before the heralding. 


To characterize the performance of the state transfer we consider the  Choi-Jamiolkowski fidelity $\mathcal{F}^{\text{CJ}}_{\text{transfer}}$ \cite{Gilchrist2005,Horodecki1999.pra.60.1888,Nielsen2002.PLA303.4}. In Sec.~\ref{sec:mainresult} we derive that in the presence of experimental imperfections of the QD-waveguide system, the infidelity is given by
\begin{align}
    1-\mathcal{F}^{\text{CJ}}_{\text{transfer}} &= \frac{3\sigma^2_o}{\Gamma^2}+\frac{3\sigma^2_e}{\Gamma^2}+\frac{3\gamma_d}{2\Gamma}
    +\frac{(\varepsilon-\gamma)^2}{2\Gamma^2}
    \nonumber\\&\quad+\frac{3\Gamma^2}{16\Delta^2}+\frac{1}{8}\frac{1}{\sigma^2_o T^{*2}_2},
    \label{eq:result1}
\end{align}
where the physical meaning of each variable is provided in Table~\ref{table:1}. Eq.~(\ref{eq:result1}) holds in the perturbative limit, corresponding to the condition set by the inequalities: $(\sigma_o,\sigma_{e},\gamma_d,\gamma,\varepsilon) \ll\Gamma \ll\Delta$ and $1/T^*_2\ll\sigma_o$. This result assumes that the protocol is  conditioned on a heralding photon in the output but no frequency filtering is assumed. The advantage  of the latter is discussed in detail below. The physical meaning of these expressions are as follows: the incident photon should have a narrow bandwidth ($\sigma_o \ll \Gamma$) and couple strongly to the waveguide ($\gamma \ll \Gamma)$ for efficient photon-emitter interaction; the spin dephasing time of the emitter needs to be longer than the duration of the incoming pulse ($T^*_2 \gg 1/\sigma_o$); the emitter should have a symmetric $\Lambda$-system ($\epsilon\ll \Gamma$) with low  coupling to its external environment $(\sigma_e,\gamma_d\ll\Gamma)$; finally $\Delta\gg\Gamma$ ensures that off-resonant spin-flip process is highly suppressed.

\setlength{\tabcolsep}{10pt}
\renewcommand{\arraystretch}{1.2}
\begin{ruledtabular}
\begin{table}[ht]
    \centering
    \begin{tabular}{cp{6.5cm}}
    Variable & Description \\ 
    \colrule
    $\Gamma$& Total decay rate of the QD (ns$^{-1}$)\\
    $\Gamma_i$& Radiative decay rate of the QD transition $\ket{e}\to\ket{g_i}$ into the waveguide (ns$^{-1}$)\\
    $\sigma_o$& Standard deviation of the spectral width of the incident Gaussian optical pulse (ns$^{-1}$)\\
    $\sigma_{e}$& Standard deviation of the optical frequency fluctuation due to spectral diffusion (ns$^{-1}$)\\
    $\gamma_d$& Pure dephasing rate (ns$^{-1}$)\\
    $\gamma$& Coupling rate to modes outside  the waveguide (ns$^{-1}$)\\ 
    $\varepsilon$& Asymmetry in decay rates: $\varepsilon=\Gamma_1-\Gamma_2$ (ns$^{-1}$) \\
    $\Delta$& Ground-state splitting: $\Delta=\omega_2-\omega_1$ (ns$^{-1}$) \\
    $T_2^*$& Spin dephasing time (ns) \\
    \end{tabular}
    \caption{Description of each variable in Eq.~(\ref{eq:result1}).}
    \label{table:1}
\end{table}
\end{ruledtabular}

Remarkably, the fidelity in Eq.~(\ref{eq:result1}) scales quadratically with almost all of the considered imperfections. Only the third term is linear. This describes pure dephasing on the optical transitions. For QDs, such dephasing arises from elastic phonon scattering that broadens the zero-phonon line~\cite{Tighineanu2018,Muljarov2004} leading to incoherent scrambling of the phase of the mapped state. In contrast inelastic phonon sidebands  can readily be removed with optical filtering and thereby be absorbed in the loss rate $\gamma$, which has a much weaker effect on the fidelity. 

The paper is structured as follows: In Sec.~\ref{sec:formalism} we revisit the formalism for photon-scattering in a waveguide, and describe the ideal state transfer scheme. In Sec.~\ref{sec:mainresult}, we evaluate how imperfections affect the state-transfer fidelity (Eq.~\eqref{eq:result1}) and give physical intuition of their roles. To explore the feasibility of the scheme, in Sec.~\ref{sec:discussion} we estimate the fidelity under realistic conditions and compare with the atom-cavity setup in Ref.~\cite{Rosenblum2017}.

\section{Photon scattering formalism}
\label{sec:formalism}
We begin this section by revisiting the photon scattering formalism in Ref.~{\cite{Witthaut2010,Das2018b}} for an emitter embedded in a one-sided waveguide. Next, we describe the ideal protocol for the photon-to-emitter state transfer.
\subsection{Photon scattering in one-sided waveguides}

We consider an emitter consisting of three levels with a $\Lambda$ configuration shown schematically in Fig.~\ref{fig:schematics}. The emitter is located at a position $z_0=0$ inside a one-sided waveguide. The Hamiltonian describing such a system ($\hbar=1$) under the rotating-wave approximation is
\begin{align}
    \mathcal{\hat{H}}&=\mathcal{\hat{H}_\text{emitter}}+\mathcal{\hat{H}_\text{field}}+\mathcal{\hat{H}_\text{int}}\nonumber\\
    &= \sum_{i=1,2} (\omega_e-\omega_i) \hat{\sigma}_{ii} + \omega_e \hat{\sigma}_{ee} -i \int dz \hat{a}^\dagger_e(z) \frac{\partial}{\partial z}\hat{a}_e(z) \nonumber\\
    &\quad + \sum_{j=1,2} \int dz \delta(z-z_0) \sqrt{\Gamma_j} \hat{a}^\dagger_e(z) \hat{\sigma}_{ej} + \text{H.c.},\label{eq:ham}
\end{align}
where $\omega_e$ is the frequency of the excited state $\ket{e}$; $\omega_i$ is the transition frequency from $\ket{g_i}\to\ket{e}$; $\hat{a}^\dagger_e$ is the photon creation operator of the even waveguided mode, which is a superposition of the right and left-propagating modes in a two-sided waveguide~\cite{Witthaut2010}, and obeys the commutator relation $[\hat{a}_e(z),\hat{a}^\dagger_e(z')]=v_g \delta(z-z')$; $v_g$ is the group velocity of the waveguided mode; $\hat{\sigma}_{ij}=\ket{j}\bra{i}$ is the atomic  operators; $\Gamma_j$ is the radiative decay rate of the transition $\ket{e}\to\ket{g_j}$ into the waveguide.

The scattering problem of a $\Lambda$-level emitter for weak input fields has been solved in Ref.~\cite{Das2018b}. From the non-Hermitian Hamiltonian $\hat{\mathcal{H}}_{nh}$ that describes the dynamics of the excited-state manifold, we can directly express the output field-mode operator of the waveguide $\hat{a}_{\text{out}}$ in terms of the incident field and dynamical response of the emitter. This allows us to easily derive the scattering matrix and spectra of the scattered fields. Assuming the emitter is initialized in the state $\ket{g_1}$, the non-Hermitian Hamiltonian describing the dynamics of the emitter is~\cite{Das2018b}
\begin{align}
    \hat{\mathcal{H}}_{\text{nh}}&=
    \bigg( \delta_1 - \frac{i\Gamma}{2}\bigg) \hat{\sigma}_{ee}\equiv\Tilde{\delta}_1 \hat{\sigma}_{ee},
    \label{eq:nh}
\end{align}
where $\delta_1 = \omega_1 - \omega$ is the detuning of the transition $\ket{g_1}\to\ket{e}$ with respect to a driving field of frequency $\omega$. 
$\Gamma=\sum_{j=1,2}  \big[\Gamma_j  + \gamma_j\big]$ is the total decay rate of the excited state $\ket{e}$ and $\gamma_j$ the radiative loss for photons emitted in the transition $\ket{e}\to\ket{j}$ that do not couple to the waveguided mode.

For a weak incident field $\hat{a}_{\text{in,e}}$ propagating in the waveguide, the output field operator $\hat{a}_{\text{out,e}}$ at an observation point $z>0$ is found to be 
\begin{align}
    \hat{a}_{\text{out,e}}(z,t) &= \bigg[1-\frac{2\Gamma_1 }{\Gamma+2i\delta_1}\hat{\sigma}_{11}  - \frac{2\sqrt{\Gamma_1 \Gamma_2}}{\Gamma+2i\delta_1}\hat{\sigma}_{12} \bigg] \nonumber\\
    &\quad\quad\times \hat{a}_{\text{in,e}}(z-v_g t).\label{eq:e1}
\end{align}
The origin of destructive interference that governs SPRINT can be inferred from Eq.~(\ref{eq:e1}):
The first and second terms represent the transmitted field and the Rayleigh-scattered field respectively. When the $\Lambda$-level system has symmetric decay rates with no loss ($\Gamma_1=\Gamma_2,\gamma=0)$ and the incident field is resonant with the transition $\ket{g_1}\to\ket{e}$ ($\delta_1=0$),
 the transmitted and scattered $\omega_1$ fields exhibit complete destructive interference, which leaves only the output at frequency  $\omega_2$ (the third term) in the waveguide and simultaneously flips the emitter's spin state.

\subsection{The ideal protocol}

In this section, we derive the ideal input-output relation (Eq.~(\ref{eq:ideal})) for the photon-to-spin state transfer based on Eq.~(\ref{eq:e1}). This input-output relation allows us to compute the state-transfer fidelity in Sec.~\ref{sec:mainresult}.

Initially, the emitter in the one-sided waveguide is in the state $\ket{g_1}$. We send a frequency-encoded qubit in the superposition state $\ket{\psi_p}_{G}=c_a \ket{\omega_1}_{G}+c_b \ket{\omega_2}_{G}$ into the waveguide, where $\ket{\omega_i}_{G}$ describes a Gaussian electric field profile $\Phi_i(\omega)$ with central frequency $\omega_i$:
\begin{align}
    \ket{\omega_i}_{G} &= \int^\infty_{-\infty} \Phi_i(\omega)\ket{\omega}d\omega \nonumber\\
    &= \int^\infty_{-\infty} (2\pi\sigma^2_o)^{-\frac{1}{4}}\exp(-\frac{(\omega-\omega_i)^2}{4\sigma^2_o})\hat{a}^\dagger(\omega)\ket{\emptyset}d\omega.
    \label{eq:gauss}
\end{align}
Note that in the definition of the input qubit state $\ket{\psi_p}_G$ we implicitly assume that the two frequency states $\ket{\omega_1}_G$ and $\ket{\omega_2}_G$ are orthogonal ${}_{G}\braket{\omega_1}{\omega_2}_G\approx0$. Due to the finite width of the Gaussian $\sigma_o$ this is a valid approximation since the overlap decreases exponentially with $\Delta/\sigma_o$ leading to an error that is negligible compared to the other infidelity terms in Eq.~(\ref{eq:result1}), which are only polynomially small. For an arbitrary frequency qubit on the Bloch sphere, $\abs{c_a}^2+\abs{c_b}^2=1$ for $c_a,c_b\in \mathbb{C}$. We then set $c_a=\cos(\theta/2)$, $c_b=e^{i\phi} \sin(\theta/2)$, where $\phi\in[0,2\pi]$ and $\theta\in[0,\pi]$. 

The frequency qubit subsequently interacts with the emitter spin in the waveguide, and the final state evolves according to
\begin{align}
    \ket{\psi_p}\otimes\ket{g_1}&=c_a \ket{\omega_1}_{G}\ket{g_1}+c_b\ket{\omega_2}_{G}\ket{g_1} \nonumber\\
    &\to\quad c_a \hat{t}^a_1 \ket{\omega_1}_{G}\ket{g_1} + c_a \hat{t}^a_2 \ket{\omega_2}_{G}\ket{g_2}\nonumber\\
    &\quad\quad + c_b \hat{t}^b_1 \ket{\omega_2}_{G}\ket{g_1} + c_b \hat{t}^b_2 \ket{\omega_2+\Delta}_{G}\ket{g_2},
    \label{eq:inputoutput}
\end{align}
where $\hat{t}^i_m$ is the scattering operator acting on the photon-spin state $\ket{\omega}_G \ket{g_m}$ for
$m\in\{1,2\}$; $i\in\{a,b\}$ indicates the type of scattering process occurred: $a$ denotes resonant scattering (SPRINT), corresponding to an input state $\ket{\omega_1}_G$, while $b$ means off-resonant scattering,  corresponding to an input state $\ket{\omega_2}_G$. We note that the scattering amplitudes should  be convoluted with the Gaussian photonic profile $\Phi(\omega)$ and integrated within the same integral.  For ease of notation we simply represent this as a frequency dependent operator  $\hat{t}^i_m$ acting on the state. $c_i$ are normalized probability amplitudes before scattering. The scattering amplitudes can be directly read from Eq.~(\ref{eq:e1}):
\begin{align}
    t^a_1 
    &= 1-\frac{2\Gamma_1}{\Gamma+2i\delta_1},\quad\quad\quad\quad\quad t^a_2 
    = \frac{-2\sqrt{\Gamma_1 \Gamma_2}}{\Gamma + 2i\delta_1}, \nonumber\\
    t^b_1
    &= 1-\frac{2\Gamma_1}{\Gamma+2i(\delta_2-\Delta)},\quad\quad t^b_2 
    = \frac{-2\sqrt{\Gamma_1 \Gamma_2}}{\Gamma + 2i(\delta_2-\Delta)},
    \label{eq:sco}
\end{align}
and are consistent with Eq.~(22) in Ref.~\cite{Witthaut2010}. $\delta_1=\omega_1 - \omega=\delta_2-\Delta$ is the laser detuning from the transition $\ket{g_1}\to\ket{e}$ for an emitter initialized in $\ket{g_1}$. 

In the ideal limit, where $\gamma=\gamma_1+\gamma_2=0$ (all emitted photons couple to the even waveguided mode), $\Gamma_1 = \Gamma_2=\Gamma/2$ (equal decay rates), $\delta_1=\delta_2=0$ (zero detuning), $\Delta\gg\Gamma$ (no off-resonant spin flip), $\sigma_o=0$ (so $\Phi_i(\omega)=\delta(\omega-\omega_i)$ for monochromatic incident field) and assuming a perfect emitter with no dephasing, the input-output relation in Eq.~(\ref{eq:inputoutput}) becomes
\begin{align}
    \ket{\psi_p}\otimes\ket{g_1}\quad
    \xrightarrow{\text{ideal}}\quad& -c_a \ket{\omega_2,g_2} + c_b \ket{\omega_2,g_1}\nonumber\\
    =&\quad\ket{\omega_2} \otimes i\hat{Y}(c_a \ket{g_1}+c_b \ket{g_2})\nonumber\\
    =&\quad\ket{\omega_2} \otimes \ket{\psi_s},\nonumber
\end{align}
which is Eq.~(\ref{eq:ideal}). It describes the ideal protocol for transferring the quantum state of an input photonic qubit $\ket{\psi_p}$ to the spin carrier $\ket{\psi_s}$, as depicted in Fig.~\ref{fig:schematics}. In this case,   perfect destructive interference between the incident and scattered fields at frequency $\omega_1$ is found $(t^a_1\to0, t^a_2\to-1$). Furthermore, $(t^b_1\to1, t^b_2\to0)$ means that the off-resonant field $\omega_2$ is fully transmitted. In practice, however,  sources of errors will affect the scattering amplitudes and thus the fidelity of the protocol. In the next section we will evaluate the influence of such errors on the fidelity.
\section{Quantum state transfer fidelity}
%
\label{sec:mainresult}
In this section, we analyze the fidelity of the quantum state transfer process. 
The general strategy that we will take is to evaluate the fidelity to lowest order in peturbation theory for each of the possible errors independently. In the end the full fidelity can then be found by adding all errors. Cross terms between different errors will only appear as higher order terms (product of errors) and thus do not appear to lowest order. This allows us to treat each imperfection separately, assuming all other parameters to have their ideal values. 

For clarity, we first divide the protocol into three different parts: resonant scattering  (Sec.~\ref{subsec:SPRINT}), off-resonant scattering (Sec.~\ref{subsec:off}), and scattering of a superposition of frequency states (Sec.~\ref{subsec:transfer}). Finally the fidelity of the whole state transfer is presented in Sec.~\ref{subsec:cjd}. 


\subsection{Resonant scattering in QD-waveguide systems}
\label{subsec:SPRINT}

When  the incident frequency qubit is $\ket{\omega_1}_G$ corresponding to $\theta=0$ or $c_a=1$, the incoming field is resonant with the QD transition $\ket{g_1}\to\ket{e}$. In this case the dynamics is given by
\begin{align}
    \ket{\omega_1}_G\ket{g_1}\quad\to\quad \hat{t}^a_1 \ket{\omega_1}_G\ket{g_1} + \hat{t}^a_2 \ket{\omega_2}_G\ket{g_2},
    \label{eq:k1out}
\end{align}
In the case of ideal SPRINT, $t_1^a=0$ and $t_2^a=-1$ results in the state $-\ket{\omega_2}_G\ket{g_2}$. Throughout this article we will consider the protocol to be conditioned on the detection of a photon in the output, since this allows us to reduce the effect of several kinds of errors. Demanding a photon in the output, we can express the fidelity of the process by
\begin{align}
    \mathcal{F}^{(c)}_{\omega_1} &= \frac{\bra{g_2}\Tr_{\omega}\rho\ket{g_2}}{\Tr\rho}\label{eq:fidelitydef}\\
    &=\frac{\eta_{\omega_2}P_{g_2}}{\eta_{\omega_1}P_{g_1}+\eta_{\omega_2}P_{g_2}}\nonumber\\
    &= 
    \begin{cases}
    \frac{P_{g_2}}{P_{g_1}+P_{g_2}} & \text{ (no filter)};\\
    1 & \text{ (filter at }\omega_2),
    \end{cases}
    \label{eq:fidelityk1}
\end{align}
where $\rho$ is the density matrix of the output state in Eq.~(\ref{eq:k1out}). In the first line the index $\omega$ on the trace indicates that we only trace over the detection of photons of frequency $\omega$. The photon detection can be frequency selective, e.g. by inserting a frequency filter before the detection of the heralding photon, so that the trace over detected photons may depend on the frequency. If a filter is present we assume the filter bandwidth to be wider than the photon bandwidth but not wider than the ground-state splitting. As such, any spectral detuning is preserved in the infidelity while different frequency components corresponding to undesired transitions are filtered out. Less than perfect filtering can  be represented by the efficiency $\eta_{\omega_i}$ of the filter at frequency $\omega_i$, but for simplicity we mainly consider no filter $\eta_{\omega_i}=1$ or perfect filtering of photons $\eta_{\omega_2}=1$, $\eta_{\omega_i}=0$ for $i\neq 2$. Since we consider an incoming photon of frequency $\omega_1$ which has the scattering amplitude $t^a_m$, the probability of the emitter to be in the state $\ket{g_m}$ after the scattering is
\begin{align}
    P_{g_m} &= \int^{\infty}_{-\infty}\abs{t^a_m(\omega) \Phi_1(\omega)}^2 d\omega,\label{eq:Pim}
\end{align}
where the integrand is the convolution of the frequency distribution of the incoming photon $\abs{\Phi_i(\omega)}^2$ with the frequency dependent scattering probability $\abs{t^a_m(\omega)}^2$. The success probability $P^s\equiv\Tr\rho$ of SPRINT is the probability of detecting a photon regardless of the final spin state:
\begin{align}
    P^s = \eta_{\omega_1}P_{g_1}+\eta_{\omega_2}P_{g_2} =     \begin{cases}
    P_{g_1}+P_{g_2} & \text{ (no filter});\\
    P_{g_2} & \text{ (filter at }\omega_2).
    \end{cases}
\end{align}

\subsubsection{Spectral mismatch errors}
The bandwidth of the pulse $\sigma_o$ described in ($\ref{eq:gauss}$) will affect the state transfer quality since the scattering probability depends on the frequency of the photons. In addition, residual broadening of the emitter may also hinder the performance via slow (compared to the emission lifetime) spectral diffusion of the QD parameterized by $\sigma_{e}$. Relevant slow spectral diffusion processes include Overhauser fluctuations from the QD nuclear spin bath \cite{Eble2009,Bulaev2005,Khaetskii2002} or electric noise from localized charged defects giving rise to Stark tuning \cite{Houel2012}. To model these effects, we introduce a small spectral shift $\delta_1\to\delta_1+\delta_e$, where $\delta_e$ follows a Gaussian spectral diffusion profile $N(0,\sigma_{e})$ with root-mean-square fluctuations $\sigma_e$.

Using~(\ref{eq:sco}),~(\ref{eq:Pim}) and the above formalism, the individual probabilities of each output state in Eq.~(\ref{eq:fidelityk1}) are
\begin{align}
    P_{g_1}
    &= \int^{\infty}_{-\infty}\frac{1}{\sqrt{2\pi\sigma^2_o}}e^{-\frac{\delta_1^2}{2\sigma_o^2}}\abs{1-\frac{\Gamma}{\Gamma+2i(\delta_1+\delta_e)}}^2 d\omega\nonumber\\
    &\approx \frac{4\sigma^2_o}{\Gamma^2} + \frac{4\delta_e^2}{\Gamma^2}=1-P_{g_2}.\label{eq:state1}
\end{align}
Averaging the fidelity $\mathcal{F}^{(c)}_{\omega_1}(\delta_e)$ over $N(0,\sigma_{e})$ results in
\begin{align}
    \mathcal{F}^{(c)}_{\omega_1}=1-\frac{4\sigma^2_o}{\Gamma^2} - \frac{4\sigma^2_e}{\Gamma^2} \quad\text{ (no filter}),
\end{align}
and $\mathcal{F}^{(c)}_{\omega_1}=1$ with filtering. The success probability is $P^s=1$ for the unfiltered case, while $P^s=1-4\sigma^2_o/\Gamma^2-4\sigma^2_e/\Gamma^2$ with filtering.
Note that for the lowest-order approximation in~(\ref{eq:state1}), the spectral fluctuations are assumed to be small compared to the QD linewidth, i.e., $\sigma_o,\sigma_{e}\ll \Gamma$. 

\subsubsection{Coupling errors}

Errors arising from imperfections in the QD-waveguide coupling include loss of the emitted photons at a rate $\gamma=\gamma_1+\gamma_2\neq0$ and asymmetric decay rates $\varepsilon=\Gamma_1-\Gamma_2\neq0$, which create an imbalance in population between basis states $\ket{\omega_2}_G \ket{g_2}$ and $\ket{\omega_2}_G \ket{g_1}$ after the state transfer~(\ref{eq:inputoutput}), reducing the transfer fidelity. The potential causes of such errors are residual coupling of the QD dipole transition to modes outside the waveguide and unequal Purcell factors of the two orthogonal dipole transitions.

To take these errors into consideration, we introduce additional loss terms in Eq.~(\ref{eq:nh}) to represent couplings to modes outside the waveguide where the total decay rate is $\Gamma=\Gamma_1+\Gamma_2+\gamma$. Including these effects yields
\begin{align}
    P_{g_1} 
    &\approx \abs{1-\frac{2\Gamma_1}{\Gamma}}^2\nonumber
    = \frac{(\varepsilon-\gamma)^2}{\Gamma^2},\\
    P_{g_2} 
    &\approx \abs{-\frac{2\sqrt{\Gamma_1\Gamma_2}}{\Gamma}}^2
    = 1 - \frac{2\gamma}{\Gamma} - \frac{(\varepsilon^2-\gamma^2)}{\Gamma^2},
\end{align}
where $\Gamma_{1/2} = (\Gamma-\gamma\pm\varepsilon)/2$. We note that since these probabilities include the probability to have a photon coming out of the waveguide they do not add up to unity when a photon is lost $\gamma\neq 0$. In particular, the transfer probability $P_{g_2}$ has a first-order dependence on the coupling loss $\gamma/\Gamma$, which reflects the probability for the scattered photon to be lost. Here we are, however, interested in the fidelity heralded on detection of a photon in the outgoing mode. In this case events where the photon is lost do not contribute to the fidelity, which is thus only affected by the probability for the system ending up in the wrong state $\ket{g_1}$ while having a photon in the waveguide. This results in the fidelity
\begin{align}
    \mathcal{F}^{(c)}_{\omega_1} &= 1 - \frac{(\varepsilon-\gamma)^2}{\Gamma^2}\quad\text{ (no filter)},
\end{align}
whereas $\mathcal{F}^{(c)}_{\omega_1}=1$ with filtering.
The reduction in fidelity is thus second order in the errors whereas the first order term only influences the success probability:
\begin{align}
    P^s &= \begin{cases}
    1-\frac{2\gamma}{\Gamma}-\frac{2\gamma(\varepsilon-\gamma)}{\Gamma^2} & \text{ (no filter)};\\
    1-\frac{2\gamma}{\Gamma}-\frac{(\varepsilon^2-\gamma^2)}{\Gamma^2} & \text{ (filter at }\omega_2).
    \end{cases}
\end{align}

\subsubsection{Phonon-induced pure dephasing}
The interaction of the QD with phonons in the semiconductor material leads to decoherence of optical transitions. This decoherence results in both broad phonon sidebands and a broadening of the zero-phonon line. The phonon sideband can be filtered away with spectral filters and its   contribution is included in the loss $\gamma$ considered above, whereas the broadening of the zero-phonon line due to elastic phonon scattering is the main source of pure dephasing in the current protocol~\cite{Besombes2001,Krummheuer2002,Muljarov2004,Tighineanu2018}.

The elastic scattering with a single phonon imprints random phases onto the excited state. We model this incoherent process as pure Markovian dephasing with a rate $\gamma_d$ and the Lindblad operator $\sqrt{2\gamma_d}\hat{\sigma}_{ee}$. In the quantum jump approach~\cite{Dalibard1992},  which is reminiscent of the approach with a non-Hermitian Hamiltonian that we consider here, this dephasing leads to a quantum jump to the excited state $\ket{e}$ followed by decay to either of the two ground states with probabilities set by the branching ratio. This leads to the density matrix
\begin{align}
    \rho' = \rho + P^{\omega_1}_{\gamma_d}\rho^{\omega_1}_{\gamma_d}\otimes\ket{g_1}\bra{g_1} + P^{\omega_2}_{\gamma_d}\rho^{\omega_2}_{\gamma_d}\otimes\ket{g_2}\bra{g_2}.\label{eq:1phonon}
\end{align}
Here $\rho$ is the density matrix in the absence of a dephasing quantum jump, and $\rho^{\omega_i}_{\gamma_d}\otimes\ket{g_i}\bra{g_i}$ are density matrices resulting from the incoherent dephasing with probabilities given by 
\begin{align}
    P^{\omega_i}_{\gamma_d}
    &= \frac{\Gamma_i}{\Gamma}P_{\gamma_d}\approx\frac{\Gamma_i}{\Gamma}\abs{\frac{-2\sqrt{2\gamma_d \Gamma^{'}_1}}{\Gamma}}^2.
    \label{eq:pured}
\end{align}

Eq.~(\ref{eq:pured}) is derived by adding a decay channel described by the dephasing operator $\hat{t}_{\gamma_d}$ to the scattering dynamics and evaluating the probability for the decay to happen through this channel. After the decay the system performs a quantum jump to the excited state (with a dephasing probability $P_{\gamma_d}$) after which it can decay to the ground state $\ket{g_i}$ with a probability $\Gamma_i/\Gamma$ by emitting a photon into the waveguide with a normalized photon density matrix $\rho^{\omega_i}_{\gamma_d}$. We here ignore any coherences between the ground states, which is justified for $\Gamma\ll \Delta$, where the emitted photons are distinguishable for the two transitions. In principle the photons emitted after the incoherent dephasing will be broader in frequency than the coherently scattered photons \cite{Childress2005} which may lead to a different filtering efficiency. For simplicity we shall, however, ignore this difference and assume the filter bandwidth to be wide enough to preserve the broadened photon bandwidth but narrow enough to filter out phonon sidebands.

In expanding $P_{\gamma_d}$ and the output state probabilities $P_{g_i}$, we substitute $\Gamma^{'}_{1/2} = (\Gamma-2\gamma_d-\gamma\pm\varepsilon)/2$ as the additional dephasing channel $\hat{t}_{\gamma_d}$ appends an imaginary term $-i\gamma_d$ to Eq.~(\ref{eq:nh})~\cite{Das2018b}. The probability for the incoherent excited state to decay is  governed only by the branching ratio $\Gamma_{1/2} = (\Gamma-\gamma\pm\varepsilon)/2$. This gives
\begin{align}
    P^{\omega_{1/2}}_{\gamma_d}&= \frac{2\gamma_d}{\Gamma}\bigg(1-\frac{2\gamma}{\Gamma}+(1\pm1)\frac{\varepsilon}{\Gamma}-\frac{2\gamma_d}{\Gamma} \bigg),\nonumber\\
    P_{g_1} &\approx \abs{1-\frac{2\Gamma^{'}_1}{\Gamma}}^2 \approx \frac{4\gamma^2_d}{\Gamma^2}-\frac{4\gamma_d(\varepsilon-\gamma)}{\Gamma^2},\nonumber\\
    P_{g_2} &\approx \abs{-\frac{2\sqrt{\Gamma^{'}_1\Gamma^{'}_2}}{\Gamma}}^2\approx 1 - \frac{4\gamma_d}{\Gamma} + \frac{4\gamma^2_d}{\Gamma^2}+\frac{4\gamma_d \gamma}{\Gamma^2}.
    \label{eq:d}
\end{align}
Eq.~(\ref{eq:d}) is simplified by considering terms only to lowest order in  $\gamma_d$ and ignoring other errors. Replacing $\rho$ in (\ref{eq:fidelitydef}) by the new spin-photon density matrix $\rho'$ (\ref{eq:1phonon}), the fidelity under phonon-induced pure dephasing is
\begin{align}
    \mathcal{F}^{(c)}_{\omega_1} &= 1-\frac{2\gamma_d}{\Gamma}\quad\text{ (no filter}),\label{eq:fd}
\end{align}
while $\mathcal{F}^{(c)}_{\omega_1}=1$ with filtering. In the latter case the infidelity is converted into inefficiency as evident from the success probability
\begin{align}
    P^s = \sum_{i=\{1,2\}} \eta_{\omega_i} (P_{g_i} + P^{\omega_i}_{\gamma_d})=\begin{cases}
    1 & \text{ (no filter)};\\
    1-\frac{2\gamma_d}{\Gamma} & \text{ (filter at }\omega_2).
    \end{cases}
\end{align}

\subsubsection{SPRINT fidelity}
Taking all the above errors into account, the SPRINT fidelity is found to be
\begin{tcolorbox}[height=1.5cm,valign=center,colback=white,boxrule=0.2mm,arc=0.2mm,top=0mm,width=8.5cm,left=0mm]
\begin{align}
    \mathcal{F}^{(c)}_{\omega_1} \approx 1- \frac{4\sigma^2_o}{\Gamma^2} - \frac{4\sigma^2_e}{\Gamma^2} - \frac{(\varepsilon-\gamma)^2}{\Gamma^2}-\frac{2\gamma_d}{\Gamma}\quad\text{(no filter)},\label{eq:flipk1}
\end{align}
\end{tcolorbox}
\noindent and $\mathcal{F}^{(c)}_{\omega_1}=1$ with filtering. The corresponding success probabilities are
\begin{align}
    P^s_{\text{unfiltered}} &=1 - \frac{2\gamma}{\Gamma}-\frac{2\gamma(\varepsilon-\gamma)}{\Gamma^2};\label{eq:eff}\\
    P^s_{\text{filtered}} &=1-\frac{4\sigma^2_o}{\Gamma^2}-\frac{4\sigma^2_e}{\Gamma^2}-\frac{2\gamma}{\Gamma}-\frac{(\varepsilon^2-\gamma^2)}{\Gamma^2}-\frac{2\gamma_d}{\Gamma}.\nonumber
\end{align}
Eq.~(\ref{eq:flipk1}) shows that 
SPRINT can be highly effective. First of all, once we filter the output, the protocol simply cannot produce wrong results for an incoming photon of frequency $\omega_1$ since the only way to produce a photon at a frequency $\omega_2$ is to decay to the desired states $\ket{g_2}$. Even without filtering SPRINT is very resilient to spectral effects like finite bandwidth $\sigma_o$ of the input photon, spectral diffusion $\sigma_e$ of the QD resonance and decay asymmetry $\varepsilon$ of the $\Lambda$ system, as all these effects enter to second order. A first order dependence is however found for the pure dephasing rate $\gamma_d$, owing to the fact that SPRINT relies on quantum interference between the incident and the scattered fields. This makes the protocol sensitive to non  phase-preserving effects which reduce this interference. While the state transfer is conditioned on detecting a photon at the output, it is noteworthy that the success probability can be arbitrarily close to unity~(\ref{eq:eff}), if all emitted photons are collected by the waveguide. This is because a photon is always emitted regardless of the branching ratio and frequency bandwidth. 
The protocol is thus near deterministic. Compared to a fully deterministic protocol not relying on the detection of a photon, the quality of the imprinted state (\ref{eq:flipk1}) is higher for the conditional protocol since we remove the dominant contribution from photon loss. 

\subsection{Off-resonant scattering in QD-waveguide systems}
\label{subsec:off}
When the incident field is in the state $\ket{\omega_2}_G$ ($\theta=\pi$), it is off-resonant from the QD transition $\ket{g_1} \to \ket{e}$ by the ground-state splitting $\Delta$. The corresponding input-output relation is
\begin{align}
    \ket{\omega_2}_G\ket{g_1}\quad\to\quad \hat{t}^b_1 \ket{\omega_2}_G\ket{g_1} + \hat{t}^b_2 \ket{\omega_2+\Delta}_G\ket{g_2}.\label{eq:off}
\end{align}
In the ideal scenario, where $\Delta\gg\Gamma$, there is no interaction between the QD and the far-detuned incident field. Hence the transmission probability approaches unity with the ideal output state $\ket{\omega_2}_G\ket{g_1}$. The fidelity conditioned on detection of a photon is therefore
\begin{align}
    \mathcal{F}^{(c)}_{\omega_2} &= \frac{\bra{g_1}\Tr_\omega\rho\ket{g_1}  }{\Tr\rho}=
    \begin{cases}
    \frac{P_{g_1}}{P_{g_1}+P_{g_2}} & \text{ (no filter)};\label{eq:flip_k3}\\
    1 & \text{ (filter at }\omega_2).
    \end{cases}
\end{align}

\subsubsection{Off-resonant Raman spin-flip error}
Unlike the first scattering process, the incident field is now highly detuned so it does not drive the QD transition. This means that the excited state will ideally not be populated.  For any finite $\Delta$ there will, however, always be a chance that the QD becomes excited and undergoes a Raman transition, resulting in the undesired final state. On the other hand, errors associated with imperfections in the QD transitions (as considered previously) become negligible compared to this dominant error since these errors will be perturbations on top of a perturbation, thus only entering to higher order. We therefore only need to evaluate the effect of having a finite $\Delta$.

The individual probabilities for the relevant output states are given by
\begin{align}
    P_{g_1} &= \int^{\infty}_{-\infty} |t^b_1 \Phi_2(\omega) |^2 d\omega\approx 1-\frac{\Gamma^2}{4\Delta^2}=1-P_{g_2},
    \label{eq:off-re}
\end{align}
where $P_{g_2}$ is the probability of undesired Raman spin-flip and $P_{g_1}$ is the transmission probability. It is apparent that without filtering the conditional fidelity for off-resonant scattering is
\begin{align}
    \boxed{\mathcal{F}^{(c)}_{\omega_2} = 1-  \frac{\Gamma^2}{4\Delta^2}\quad\text{ (no filter),} }\label{eq:flipk3}
\end{align}
with $P^s=1$. Similarly we again have $\mathcal{F}^{(c)}_{\omega_2}=1$ for the filtered case since a decay to the wrong state will always be associated with a frequency shift. The corresponding success probability for the filtered case is $1-  \Gamma^2/4\Delta^2$. 
It is important to emphasize that the off-resonant spin-flip error $\Gamma^2/4\Delta^2$ is different from the error coming from the frequency overlap of the two incoming states $\ket{\omega_1}_G$ and $\ket{\omega_2}_G$. Assuming a Gaussian input field, the outgoing scattered field will also be a Gaussian with equal spectral width $\sigma_o$ as the input. This overlap error scales exponentially ${}_G\braket{\omega_1}{\omega_2}_G=\exp(-\Delta^2/8\sigma^2_o)$. Since a high fidelity for both scattering cases require $\sigma_o^2\ll\Gamma^2\ll\Delta^2$, this error is negligible compared to the terms we consider. 
%
%
\subsection{State transfer of an equatorial photonic state to a QD spin}
\label{subsec:transfer}
So far we have considered the extreme cases where the photon is initialized in one of the two frequency states. For a more general case, we here consider a state in which the incoming photon is in an equal superposition of two frequencies $(\ket{\omega_1}_G+e^{i\phi}\ket{\omega_2}_G)/\sqrt{2}$ (corresponding to $\theta=\pi/2)$. The spin-photon scattering process is then described by Eq.~(\ref{eq:inputoutput}):
\begin{align}
    &\frac{1}{\sqrt{2}}\bigg(\ket{\omega_1}_G +e^{i\phi}\ket{\omega_2}_G\bigg) \otimes \ket{g_1} 
    \nonumber\\
    &\quad\to \frac{1}{\sqrt{2}}\bigg(\hat{t}^a_1 \ket{\omega_1}_{G}\ket{g_1} +  \hat{t}^a_2 \ket{\omega_2}_{G}\ket{g_2} \nonumber\\
    &\quad\quad + e^{i\phi}\hat{t}^b_1\ket{\omega_2}_{G}\ket{g_1} + e^{i\phi}\hat{t}^b_2 \ket{\omega_2+\Delta}_{G}\ket{g_2}\bigg)=\ket{\Psi}.
    \label{eq:transfer}
\end{align}
Here $\ket{\Psi}$ denotes the spin-photon output state. The conditional fidelity of the produced state is then 
\begin{align}
    \mathcal{F}^{(c)}_{\phi} = \bra{\Psi^{s}_{\text{ideal}}}\rho^{(s)}\ket{\Psi^{s}_{\text{ideal}}},\label{eq:eqt}
\end{align}
where $\ket{\Psi^s_{\text{ideal}}}=i\hat{Y}(\ket{g_1}+e^{i\phi}\ket{g_2})/\sqrt{2}$ is the ideal spin state. Here $\rho^{(s)}$ is the reduced density matrix of the spin system given by partial trace of the output density matrix $\ket{\Psi}\bra{\Psi}$ 
over the frequency of the outgoing photon in the waveguide
\begin{align}
    \rho^{(s)} =\frac{\Tr_{\omega}(\ket{\Psi}\bra{\Psi}\hat{P}_{\omega})}{\Tr(\ket{\Psi}\bra{\Psi}\hat{P}_{\omega})},\label{eq:rdm}
\end{align}
where $\hat{P}_{\omega}$ is a projection operator representing  filtering: For a filter at $\omega_j$, $\hat{P}_{\omega}\ket{\omega_i}=\delta_{ij}\ket{\omega_i}$; if there is no filter, the projector is just  the identity operator $\hat{P}_{\omega}=\mathcal{I}$. The denominator in (\ref{eq:rdm}) gives the success probability $P^s$.

\subsubsection{Spin dephasing error}
For a faithful photon-to-spin state transfer, it is essential that the phase of the photonic qubit is preserved in the mapped spin state. This means the spin qubit must remain coherent before being read out. However, the QD spin is coupled to a neighboring nuclear spin bath via hyperfine interaction, causing the spin to precess with a fluctuating frequency $\delta_g$. This is referred to as the Overhauser noise \cite{Eble2009,Bulaev2005,Khaetskii2002}, which is one of the external sources that limit the spin coherence time. This dephasing effect can be  modelled by slow drifts in the energy levels of the $\Lambda$-system. 
\begin{figure}[ht]
\includegraphics[width=1\linewidth]{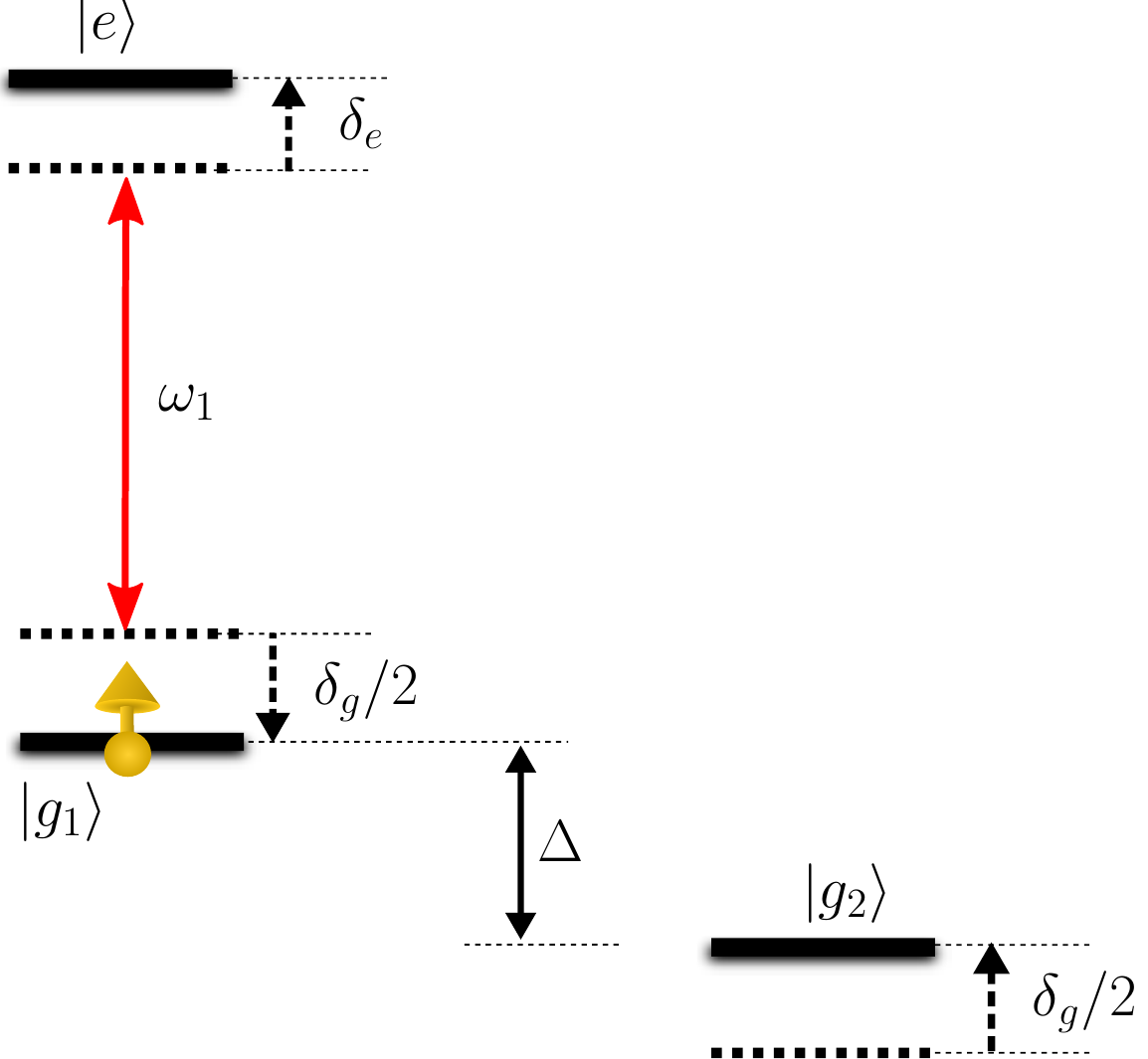}
\caption{Energy levels of the $\Lambda$-system under spin dephasing noise and spectral diffusion.}
\label{fig:corr}
\end{figure}
As illustrated in Fig.~\ref{fig:corr} the ground-state dephasing effectively detunes the energy levels, shifting the QD resonance such that $\omega_1\to\omega_1+\delta_g/2$, similar to the frequency shift $\delta_e$ due to spectral diffusion considered above. 

Now we first evaluate the fidelity for the unfiltered case. The ground-state populations after state transfer are given by
\begin{align}
    \rho^{(s)}_{11} &= \frac{1}{2P^s}\bigg[ \int^{\infty}_{-\infty}\abs{t^a_1\Phi_1(\omega)}^2 d\omega + \int^{\infty}_{-\infty}\abs{t^b_1\Phi_1(\omega)}^2 d\omega\bigg]\nonumber\\
    &=\frac{1}{2P^s}\bigg(1+\frac{4\sigma^2_o}{\Gamma^2}+\frac{\delta^2_g}{\Gamma^2}\bigg) = 1-\rho^{(s)}_{22},\label{eq:gp}
\end{align}
where we assume $\Delta\gg\delta_g$ for small fluctuations. The success probability for the unfiltered case is unity  $P^s=\rho^{(s)}_{11}+\rho^{(s)}_{22}=1$ in the absence of coupling loss. For now we ignore the dephasing of the excited state $\delta_e\to0$ and focus only on the ground-state coherence. In general the Overhauser field and other noise sources will shift  $\delta_e$ and $\delta_g$ in a possibly  correlated manner depending on the system investigated. For simplicity we assume the fluctuations to be uncorrelated with independent Gaussian noise distributions. This means the linear cross error $\delta_e\delta_g$ will be averaged out, leaving only second-order dependence on $\delta_e$ in the final fidelity. This will be considered below and for now we focus on the ground state coherence.

For the off-diagonal spin coherence terms we find
\begin{align}
    \rho^{(s)}_{12}&=\frac{1}{2P^s}\int^{\infty}_{-\infty}t^{a*}_2 t^b_1 e^{i\phi}\abs{\Phi_1\Phi_2}d\omega\nonumber\\
    &\approx \frac{-e^{i\phi}}{2P^s}\bigg(1-\frac{4\sigma^2_o}{\Gamma^2}-\frac{\delta^2_g}{\Gamma^2}+i\frac{\delta_g}{\Gamma}\bigg)=\rho^{(s)*}_{21}.
\end{align}
Note that $\rho^{(s)}$ is time-independent as it has been derived in a rotating frame with respect to a fixed ground-state splitting $\Delta+\delta_g$ (Fig.~\ref{fig:corr}). Since the ground-state splitting may vary from shot to shot of the experiments, the rotating frame now depends on the splitting and we need to take this variation into account. To do this we assume the evolution before the emission of the photon is independent of the splitting since there is no superposition of the ground states before this time. Here we consider heralded operation where the photon is detected in the output. It is therefore convenient to analyze what happens for a specific photon detection time.  

Let $t_c$ be the time of detecting a photon in the output (or equivalently the creation time of the spin qubit). The spin precession induced by the Overhauser noise is then denoted by the time evolution operator $\hat{T}(t-t_c)=\exp(-i\delta_g \hat{S}_z (t-t_c))$, with $\hat{S}_z=\hat{\sigma}_z/2$. 

Since the protocol creates a superposition of two ground states which are subject to Overhauser noise, it is beneficial to incorporate a spin-echo sequence into the protocol to extend the spin coherence time.  In the echo sequence a $\pi$ pulse is applied at some time $t_\pi$. Ideally the time interval between the detection of the photon and the $\pi$-pulse will be equal to the time interval between the $\pi$-pulse and the final readout time $t_R$, i.e. $t_R-t_\pi=t_\pi-t_c=T$. This ensures complete refocusing of  the spin state at $t_R$ such that any effect of the spin decoherence is removed~\cite{Hahn1950,Koppens2008,Wang2012}. 

\begin{figure}[ht]
\includegraphics[width=1\linewidth]{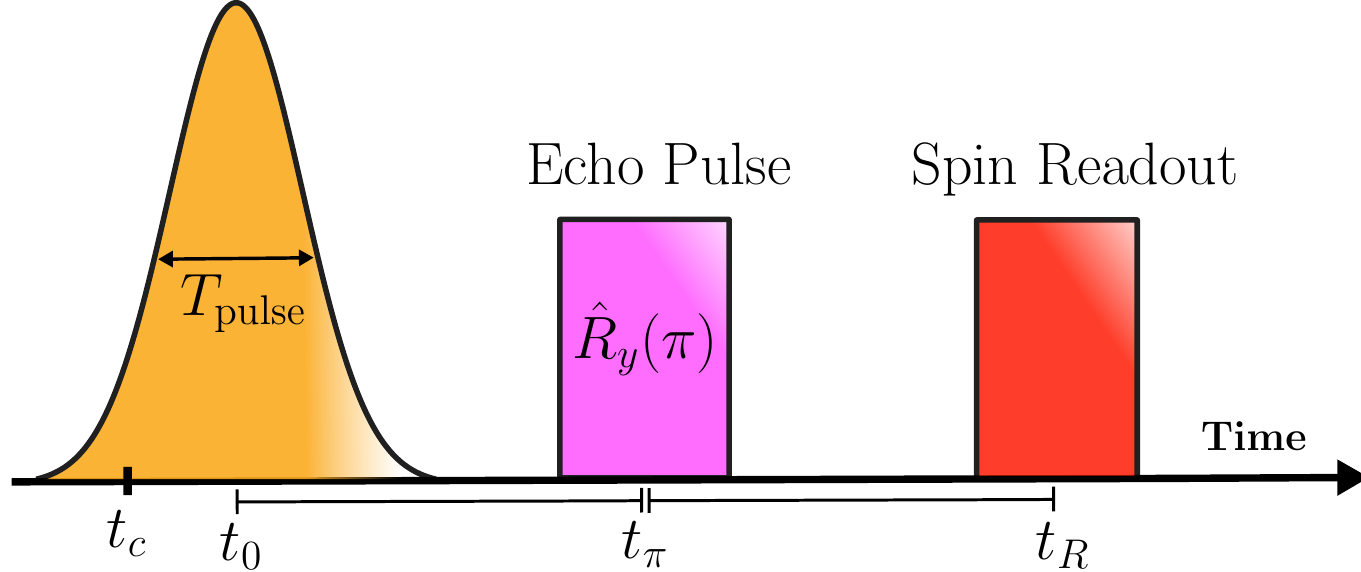}
\caption{Pulse sequence with spin echo. The spin echo sequence assumes that the state transfer occurs at $t_0$, and  the $\pi$-pulse and spin readout are applied at $t_\pi$ and $t_R$ respectively such that $t_R - t_\pi = t_\pi - t_0$. $t_c$ follows the probability distribution of the scattered pulse.}
\label{fig:echo}
\end{figure}

In practice, however, the  time of the photon detection $t_c$ cannot be determined exactly due to time jittering of the photodetector, and it is complicated to make the time of the readout or $\pi$-pulse be dependent on the detection time. For this reason we consider a simpler experimental procedure, where the spin echo sequence begins at a predetermined time $t_0$ given by the peak of the outgoing pulse (Fig.~\ref{fig:echo}). Since in general this implies $t_c\neq t_0$, $t_R-t_\pi\neq t_\pi - t_c$ this means the echo becomes imperfect, leading to dephasing of the spin qubit.

The evolution of the spin system under spin echo is described by the unitary operator 
\begin{align}
    \hat{\mathcal{U}}_{\text{echo}} = \hat{T}(t_R-t_\pi)\hat{R}_y(\pi)\hat{T}(t_\pi-t_c),\label{eq:ses}
\end{align}
where $\hat{R}_y(\pi)=e^{i\pi\hat{\sigma}_y/2}$ is the spin echo $\pi$-pulse applied at $t_\pi$. 
Using Eq. (\ref{eq:ses}) and the echo condition $t_R-t_\pi=t_\pi-t_0=T$, the spin reduced density matrix becomes
\begin{align}
    \rho^{(s)}_{\text{echo}} &= \hat{\mathcal{U}}_{\text{echo}} \rho^{(s)} \hat{\mathcal{U}}_{\text{echo}}^\dagger\nonumber\\
    &= \left[
    \begin{array}{cc}
        \rho^{(s)}_{22} & -\rho^{(s)}_{21}e^{-i\delta_{g}(t_c-t_0)}\\ 
        -\rho^{(s)}_{12}e^{i\delta_{g}(t_c-t_0)} & \rho^{(s)}_{11}
            \end{array}
             \right].\label{eq:srdm}
\end{align}
From Eq.~(\ref{eq:srdm}) one can infer that for perfect spin echo ($t_c=t_0$) the coherence of the mapped spin state is preserved. Using Eq. (\ref{eq:eqt}), the conditional fidelity is 
\begin{align}
    &\mathcal{F}^{(c)}_{\phi}(\delta_g,t_c)= \bra{\Psi^s_{\text{ideal}}}\hat{R}^\dagger_y(\pi)\rho^{(s)}_{\text{echo}}\hat{R}_y(\pi)\ket{\Psi^s_{\text{ideal}}}\nonumber\\
    &=\frac{1}{2P^s}\bigg[1+\bigg(1-\frac{4\sigma^2_o}{\Gamma^2}-\frac{\delta^2_g}{\Gamma^2} +i\frac{\delta_g}{\Gamma}\bigg)\cos(\delta_g (t_c-t_0))\bigg],\label{eq:eqtb}
\end{align}
where the ideal spin state now becomes $\hat{R}_y(\pi)\ket{\Psi^s_{\text{ideal}}}$ due to the $\pi$-rotation. The mean fidelity is  calculated by averaging (\ref{eq:eqtb}) with respect to the noise distributions. Averaging over the dephasing profile $N(0,\sigma_g)$ with $\sigma_g=\sqrt{2}/T^*_2$ gives the (unfiltered) fidelity
\begin{align}
    &\bar{\mathcal{F}}^{(c)}_{\phi}(t_c)
    =\int^{\infty}_{-\infty}\int^{\infty}_{-\infty}\mathcal{F}^{(c)}_{\phi}(\delta_g,t_c,\delta_e)N(0,\sigma_g) d\delta_g\nonumber\\
    &=\frac{1}{2P^s}\bigg[1+e^{-\frac{1}{2}\sigma^2_g (t_c-t_0)^2} \bigg(1-\frac{4\sigma^2_o}{\Gamma^2}-\frac{\sigma^2_g}{\Gamma^2}\bigg)\bigg].
    \label{eq:bjitter}
\end{align}

Recall that the detector click occurs at $t_c$ within the time window of the scattering pulse, and the echo starts at $t_0$. 
Since we do not condition on the exact time of the detector click we need to average over the probability distribution of the outgoing photons. In principle, due to scattering this distribution deviates from the distribution of the incoming photon. In order to minimize the error coming from the bandwidth of the pulse $\sim\sigma_o^2/\Gamma^2$, we are interested in the regime $T_{{\rm pulse}}\gg  1/\Gamma$ where $T_{\text{pulse}}=1/(2\sigma_o)$ is the transform-limited pulse duration.
In this regime, the temporal profile of the scattering pulse is only slightly perturbed thus the probability distribution of the outgoing pulse can be approximated by the Gaussian profile $N(t_0,T_{\text{pulse}})$ of the incoming pulse. Averaging Eq.~(\ref{eq:bjitter}) over $N(t_0,T_{\text{pulse}})$ gives
\begin{align}
    \bar{\mathcal{F}}^{(c)}_{\phi}&=\int^{\infty}_{-\infty} \bar{\mathcal{F}}^{(c)}_{\phi}(t_c)N(t_0,T_{\text{pulse}})dt_c\nonumber\\
    &=\frac{1}{2P^s}\bigg[1+\frac{1}{\sqrt{1+\sigma^2_g T^2_{\text{pulse}}}}\bigg(1-\frac{4\sigma^2_o}{\Gamma^2}-\frac{\sigma^2_g}{\Gamma^2} \bigg)\bigg]\nonumber\\
    &\approx 1-\frac{2\sigma^2_o}{\Gamma^2}-\frac{1}{8}\frac{1}{\sigma^2_o T^{*2}_2}-\frac{1}{ T^{*2}_2 \Gamma^2}\quad\text{ (no filter}).\label{eq:eqtr}
\end{align}
From here we see there is an optimal bandwidth $\sigma_o=2\sqrt{\Gamma/T_2^*}$ for the incident photon as dictated by the trade-off between the first and second error terms: For incident photons with larger bandwidth $\sigma_o$ (or shorter $T_{\text{pulse}}$),
the frequency mismatch between the photon and the QD linewidth $\Gamma$ is higher, which lowers the probability of driving the Raman transition, and eventually the state-transfer fidelity. On the other hand, for narrow photons (or long $T_{\text{pulse}}$), the transfer becomes more prone to spin dephasing. The reason for this is that  a longer transfer duration $T_{\text{pulse}}$ implies a larger uncertainty in the creation time $t_c$ of the spin qubit, which renders the spin echo ineffective. We therefore require $T_{\text{pulse}}\ll T^*_2$ in order for  the spin qubit to remain coherent.

The third infidelity term in Eq. (\ref{eq:eqtr}) originates from the energy shift in the ground states due to the Overhauser noise, which shifts the QD resonance similar to spectral diffusion. This error is always smaller than the combination of the first and third terms since these require $1/{T_2^*}^2\ll \sigma_o^2\ll \Gamma^2$. Furthermore experimental values confirm that this term is small for realistic systems, e.g. $T^*_2 \Gamma \approx50$ \cite{Appel2020}. 

In the case of filtering photons at the frequency $\omega_2$, the spin populations (\ref{eq:gp}) are reduced due to the filter. Apart from that the calculation proceeds along the same lines and leads to a lower success probability but enhanced fidelity
\begin{align}
    P^s &= \frac{1}{2} \int\bigg(\abs{t^a_2}^2+ \abs{t^b_1}^2 \bigg)\abs{\Phi_1(\omega)}^2d\omega\approx1-\frac{2\sigma^2_o}{\Gamma^2},\nonumber\\
    \bar{\mathcal{F}}^{(c)}_{\phi} &\approx 1-\frac{\sigma^2_o}{\Gamma^2}-\frac{1}{8}\frac{1}{\sigma^2_o T^{*2}_2}\quad\text{ (filter at }\omega_2).
\end{align}
\subsubsection{Equatorial fidelity}
We conclude this section by including all the relevant errors in addition to spin dephasing. The spectral mismatch, coupling and off-resonant spin-flip infidelities are calculated straightforwardly similar to Secs.~\ref{subsec:SPRINT} and \ref{subsec:off}, whereas the phonon-induced pure dephasing transforms the output density matrix $\rho=\ket{\Psi}\bra{\Psi}$ into
\begin{align}
    \rho' = \rho + \frac{1}{2}\bigg(P^{\omega_1}_{\gamma_d}\rho^{\omega_1}_{\gamma_d}\otimes\ket{g_1}\bra{g_1} + P^{\omega_2}_{\gamma_d}\rho^{\omega_2}_{\gamma_d}\otimes\ket{g_2}\bra{g_2}\bigg),\nonumber
\end{align}
since only half of the incoming qubit resonantly drives the QD to the excited state. As such, the effect from phonon scattering is halved compared to Eq.~(\ref{eq:1phonon}).

Including all effects, the corresponding fidelity is
\begin{tcolorbox}[height=4cm,valign=center,colback=white,boxrule=0.2mm,arc=0mm,top=0mm,width=8.5cm,left=0mm]
\begin{numcases}{\mathcal{\bar{F}}^{(c)}_{\phi}=}
    1-\frac{2\sigma^2_o}{\Gamma^2}-\frac{2\sigma^2_e}{\Gamma^2}-\frac{\gamma_d}{\Gamma}
    -\frac{(\varepsilon-\gamma)^2}{4\Gamma^2}\nonumber\\
    \quad\quad-\frac{1}{8}\frac{1}{\sigma^2_o T^{*2}_2}-\frac{\Gamma^2}{8\Delta^2}\quad\quad\text{(no filter);}\nonumber\\
    1-\frac{\sigma^2_o}{\Gamma^2}-\frac{\sigma^2_e}{\Gamma^2}-\frac{\gamma_d}{2\Gamma}
    -\frac{1}{8}\frac{1}{\sigma^2_o T^{*2}_2}\\
    \quad\quad-\frac{\Gamma^2}{16\Delta^2}\quad\quad\quad\quad\text{ (filtered at }\omega_2)\nonumber,
\end{numcases}
\end{tcolorbox}
\noindent with the success probability 
\begin{align}
    P^s_{\text{unfiltered}} &=1-\frac{\gamma}{\Gamma}-\frac{\gamma(\varepsilon-\gamma)}{\Gamma^2};\\
    P^s_{\text{filtered}}&=1-\frac{2\sigma^2_o}{\Gamma^2}-\frac{2\sigma^2_e}{\Gamma^2}-\frac{\gamma}{\Gamma}-\frac{(\varepsilon^2-\gamma^2)}{2\Gamma^2}-\frac{\gamma_d}{\Gamma}-\frac{\Gamma^2}{8\Delta^2}.\nonumber
\end{align}
\subsection{Choi-Jamiolkowski fidelity}
\label{subsec:cjd}
Above we have characterized the performance of the state transfer for specific states. 
To quantify the overall performance for an unknown state, it is however, crucial to average the transfer fidelity over different input states on the Bloch sphere. The Choi-Jamiolkowski fidelity $\mathcal{F}^{\text{CJ}}$ is an example of such a metric \cite{Gilchrist2005,Horodecki1999.pra.60.1888,Nielsen2002.PLA303.4} which is useful for characterizing noisy quantum gates and channels. For instance $\mathcal{F}^{\text{CJ}}\geq50\%$ guarantees that an operation is entanglement-preserving. 

In essence, if $\mathcal{E}$ denotes a noisy  operation acting on the spin-photon system $S$, the input state $\ket{\psi_{\text{in}}}$ is thought to be a maximally entangled state between $S$ and an ancillary system $A$. Then the Choi-Jamilkowski fidelity $\mathcal{F}^{\text{CJ}}\equiv\bra{\psi_{\text{ideal}}}(\mathcal{I}_A\otimes\mathcal{E}_S)(\ket{\psi_{\text{in}}}\bra{\psi_{\text{in}}})\ket{\psi_{\text{ideal}}}$ measures the fidelity of  the entangled state  after applying $\mathcal{E}$ to half of the input state, where $\mathcal{I}_A$ is the identity operation on system $A$ and $\ket{\psi_{\text{ideal}}}$ is the ideal state.

In our case we consider a gate operation heralded on the detection of a photon in the output mode. The success probability for the gate depends on which input state we consider and this needs to be taken into account when assessing its performance. In Appendix A we generalize the formula for average gate fidelity to heralded operation in a qubit system~\cite{Nielsen2002.PLA303.4}. Specifically we show that for non-trace preserving gate operations $\mathcal{E}$ (i.e., our heralded state-transfer protocol), there is a linear relationship between the Choi-Jamiolkowski fidelity and the weighted average fidelity $\bar{\mathcal{F}}^{(c)}_{\text{weighted}}$:
\begin{align}
    \boxed{\bar{\mathcal{F}}^{(c)}_{\text{weighted}}\equiv\frac{\sum_i P^s_i \mathcal{F}^{(c)}_i}{\sum_i P^s_i}=\frac{2}{3}\mathcal{F}^{\text{CJ}}+\frac{1}{3},}
    \label{eq:linear}
\end{align}
where the index $i$ refers to different input photonic states on $6$ cardinal points of the Bloch sphere: $i\in\{1,2\}$ correspond to conditional fidelities of SPRINT and off-resonant scattering in Eqs. (\ref{eq:flipk1}) and~(\ref{eq:flipk3}) respectively, whereas $i\in[3,6]$ represent fidelities for $4$ superposition states on the equator of the Bloch sphere. $P^s_i$ is the success probability for each transfer characterized by the number of detected photons.

Eq.~(\ref{eq:linear}) enables us to extract $\mathcal{F}^{\text{CJ}}$ by covering only $6$ cardinal input states on the Bloch sphere and measuring their conditional fidelities. Here we use this formula to evaluate the Choi-Jamilkowski fidelity for our protocol, but this also
represents a receipe for how to construct a suitable average in experiments. For instance the performance of single qubit storage is often evaluated by averaging the fidelity over the Bloch sphere and demanding that the fidelity exceeds a classical threshold of 2/3 \cite{Massar1995,Hammerer2005}. Our result indicates that this condition is equivalent to having an entanglement-preserving operation  $\mathcal{F}^{\text{CJ}}\geq 1/2$,  if the experimental results are weighed with the success probability as in Eq. (\ref{eq:linear}). 

We note that Eq. ($\ref{eq:linear}$) can be used to evaluate the averaged fidelity which suits different experimental scenarios: Heralding on the arrival of photons with or without a filter at $\omega_2$ result in
\begin{tcolorbox}[height=3.2cm,valign=center,colback=white,boxrule=0.2mm,arc=0mm,top=0mm,width=8.5cm,left=0mm]
\begin{align}
    \mathcal{F}^{\text{CJ}}_{\text{unfiltered}}
    &=1-\frac{3\sigma^2_o}{\Gamma^2}-\frac{3\sigma^2_e}{\Gamma^2}-\frac{3\gamma_d}{2\Gamma}
    -\frac{(\varepsilon-\gamma)^2}{2\Gamma^2}
    \nonumber\\
    &\quad\quad-\frac{1}{8}\frac{1}{\sigma^2_o T^{*2}_2}-\frac{3\Gamma^2}{16\Delta^2},\label{eq:3def}\\
    \mathcal{F}^{\text{CJ}}_{\text{filtered}}
    &=1-\frac{\sigma^2_o}{\Gamma^2}-\frac{\sigma^2_e}{\Gamma^2}-\frac{\gamma_d}{2\Gamma}-\frac{1}{8}\frac{1}{\sigma^2_o T^{*2}_2}-\frac{\Gamma^2}{16\Delta^2}.\nonumber
\end{align}
\end{tcolorbox}
\begin{figure}
	\includegraphics[width=1\linewidth]{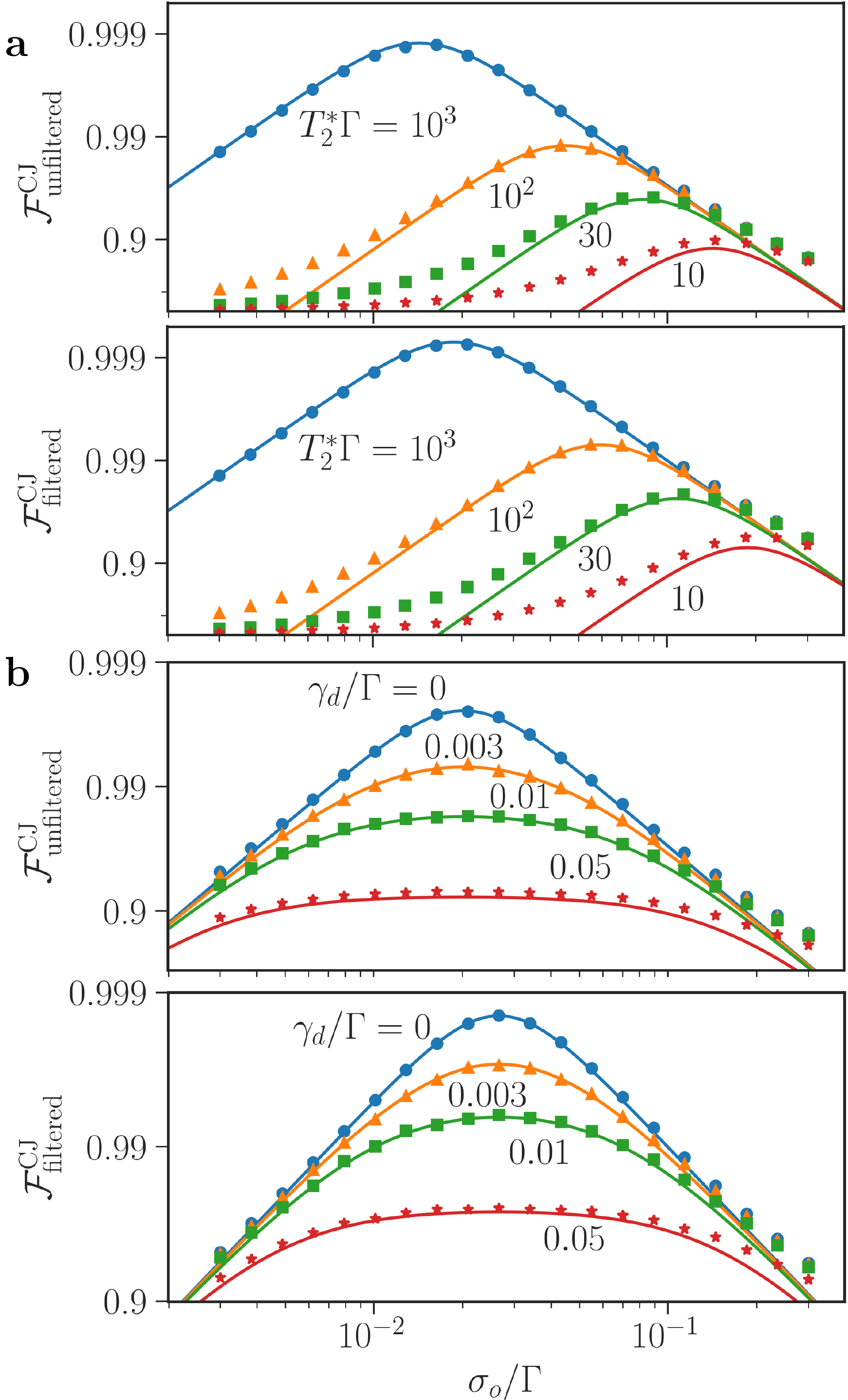}
	\caption{(color online) Choi-Jamiolkowski fidelities of the photon-to-spin state transfer with/without filtering as a function of the bandwidth of the input field $\sigma_o$. Solid lines correspond to theoretical values from Eq.~(\ref{eq:3def}). Symbols are simulated results. \textbf{(a)} Varying spin dephasing time $T_{2}^{*}$. All other errors are ignored, i.e. $\gamma_d=0$, $\varepsilon=0$, $\sigma_e=0$,  $\Gamma/\Delta\approx0$ and  $\gamma=0$. \textbf{(b)} Varying pure dephasing rate $\gamma_d$ with $ T_{2}^{*}\Gamma=500$. $\varepsilon=0$, $\sigma_e=0$,  $\Gamma/\Delta\approx0$ and  $\gamma=0$.}
	\label{fig:sigma_dep}
\end{figure}
\begin{figure}
	\includegraphics[width=1\linewidth]{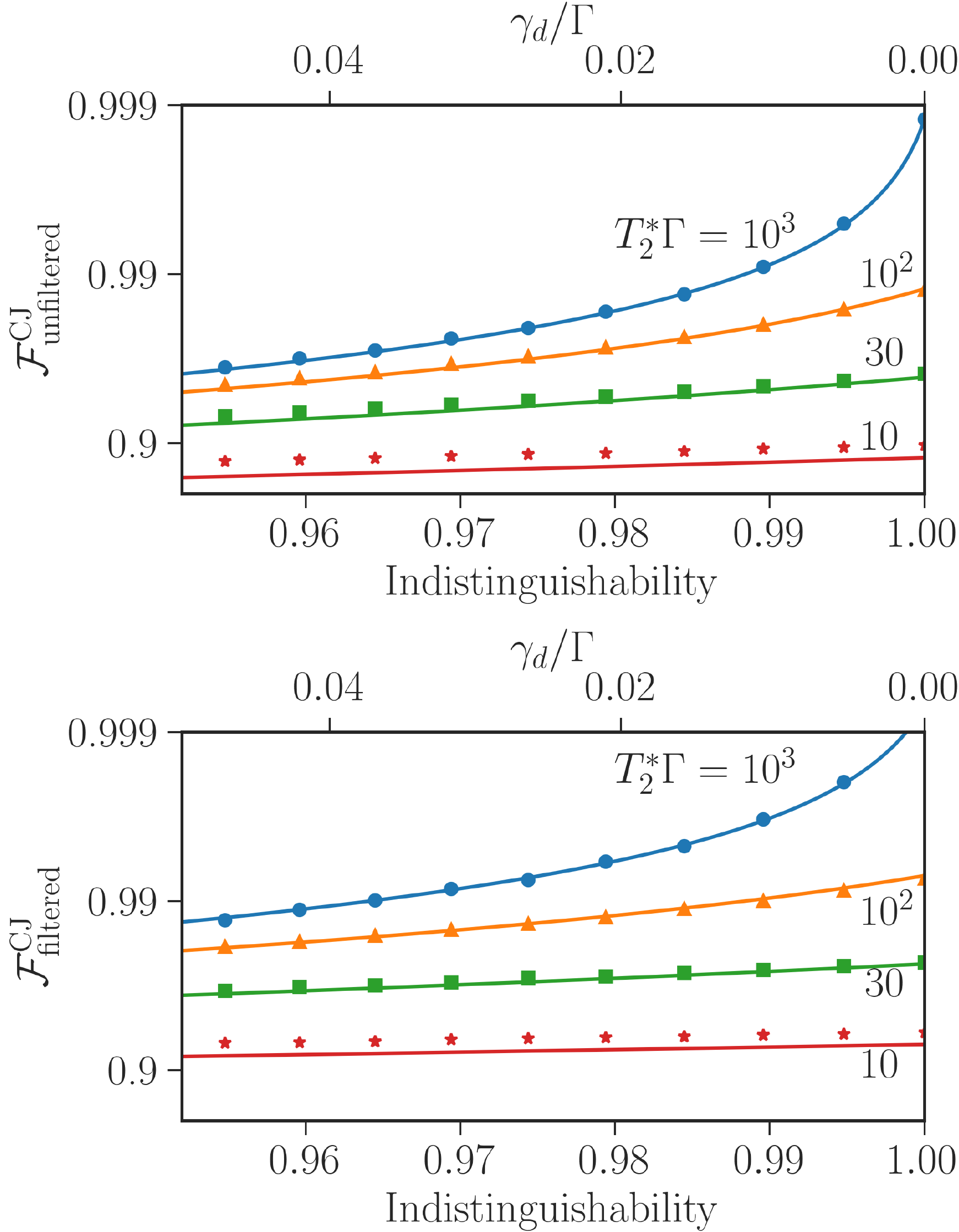}
    \caption{(color online) Choi-Jamiolkowski fidelities with/without filtering as a function of the dephasing rate $\gamma_d$ or equivalently the HOM indistinguishability $I = \Gamma / (\Gamma + 2\gamma_d)$ for various spin dephasing times $T^{*}_2$ assuming optimal bandwidth of the input field. Solid lines correspond to the analytical result in Eq.~(\ref{eq:3def}). Symbols are simulated results. All other errors are neglected, i.e. $\varepsilon=0$, $\sigma_e=0$,  $\Gamma/\Delta\approx0$ and $\gamma=0$.}
	\label{fig:I}
\end{figure}
The corresponding success probabilities are computed as a direct average over the Bloch sphere:
\begin{align}
    P_{\text{unfiltered}}^s &= 1-\frac{\gamma}{\Gamma}-\frac{\gamma(\varepsilon-\gamma)}{\Gamma^2},\label{eq:P}\\
    P_{\text{filtered}}^s &=1-\frac{2\sigma^2_o}{\Gamma^2}-\frac{2\sigma^2_e}{\Gamma^2}-\frac{\gamma}{\Gamma}-\frac{(\varepsilon^2-\gamma^2)}{2\Gamma^2}-\frac{\gamma_d}{\Gamma}-\frac{\Gamma^2}{8\Delta^2}\nonumber
    .
\end{align}
Here we again see that the filter increases the fidelity although at the expense of lowering the success probability of state transfer.

So far all results have been derived analytically with perturbation theory in order to obtain a solid understanding of the errors. To verify our results and to be able to go beyond perturbation theory we have also simulated the protocol numerically. For this purpose, we model the interaction between an incident photon and a  waveguide-embedded QD using the quantum trajectory theory of cascaded open systems \cite{carmichael1993quantum}. The evolution of the state according to the effective non-Hermitian Hamiltonian is computed by applying the Monte-Carlo wave-function procedure \cite{Dalibard1992,molmer1993monte,carmichael1993quantumlec} with over $10^5$ stochastic wave-functions per simulated point. The scattering probabilities are then extracted from the simulated wave-functions to evaluate the averaged fidelities. Figs. \ref{fig:sigma_dep} and \ref{fig:I} show an excellent agreement between the analytical and numerical results in the limit of $1-\mathcal{F}^{\text{CJ}}\ll 1$, which is the regime where we expect perturbation theory to hold.



\section{Experimental considerations}
\label{sec:discussion}

We now discuss the possible implementation of the photon-to-spin state transfer using a QD coupled to nanophotonic waveguides. In particular, we estimate the maximum attainable fidelity using experimentally demonstrated parameters. We then compare  the performance of a waveguide-QD system to that of a single atom coupled to an optical cavity, a platform previously used to demonstrate a quantum SWAP gate~\cite{Rosenblum2017,Bechler2018}. 

A QD coupled to a photonic crystal waveguide involving a positive (XP) or negative (XM) charge under a Voigt magnetic field could be used to realize the optical $\Lambda$-level system. So far we have considered a one-sided (i.e., terminated) waveguide, while in principle a two-sided configuration is also feasible. In that case, however, optimal performance requires excitation and collection from both sides of the waveguide with mutual interferometric stability, which would be an experimental overhead to implement~\cite{Witthaut2012}. As such, we here consider a one-sided waveguide.

For strong radiative coupling, the essential efficiencies include the ratio of the emission into the coherent zero-phonon line where an efficiency of $95\%$ has been reported~\cite{Borri2005,Hansom2014}, and the coupling efficiency into the waveguided mode which has been found to be exceedingly high with an efficiency ($\beta$-factor) of 98\%~\cite{Arcari2014}. These two efficiencies indicate that $\gamma/\Gamma<7\%$ is achievable, only limiting $\mathcal{F}^{\text{CJ}}_{\text{unfiltered}}$ and $P_{\text{unfiltered}}^s$ to $99.5\%$ and $93\%$, respectively.

To minimize spin-related errors, we consider an optically excited positively charged exciton state (XP). Hole spins are proven to have significantly longer spin dephasing times $T^*_2$ than electrons~\cite{Warburton:2013aa} without additional cooling of the nuclear spin ensemble, and are thus  suitable for our protocol. Recently, $T_2^*=21.4$~ns was reported for a hole spin in a photonic crystal waveguide along with a XP radiative decay rate of $\Gamma=2.48$~ns$^{-1}$~\cite{Appel2020}. This translates into $T_2^*\Gamma\approx 54$ and a corresponding $\mathcal{F}^{\text{CJ}}_{\text{unfiltered}}$ of 97.7\% when the photon bandwidth is optimally engineered ($\sigma_{o,\text{optimal}}=24^{-\frac{1}{4}}\sqrt{\Gamma/T^*_2}\approx0.15 \text{ ns}^{-1}$). 

As noted previously, the most prominent infidelity results from the linear scaling with the pure dephasing rate $\gamma_d$. The pure dephasing rate is directly measured in a Hong-Ou-Mandel (HOM) interference experiment where the degree of indistinguishability between two emitted photons is recorded and expressed as $I = \Gamma / (\Gamma + 2\gamma_d)$. $I>98\%$ (corrected for two-photon components) has been reported for two subsequently emitted photons \cite{Ding2016}, which translates into $\gamma_d/\Gamma \approx 1\%$ corresponding to a maximum $\mathcal{F}^{\text{CJ}}_{\text{unfiltered}}$ of 98.5\%.

The decay rate asymmetry $\varepsilon$ could in principle be reduced by rotating the in-plane external magnetic field or by properly positioning the QD in the photonic crystal waveguide. Furthermore, this error only enters to the second order and is not expected to pose a fundamental limitation. 

\begin{figure}
    \includegraphics[width=1\linewidth]{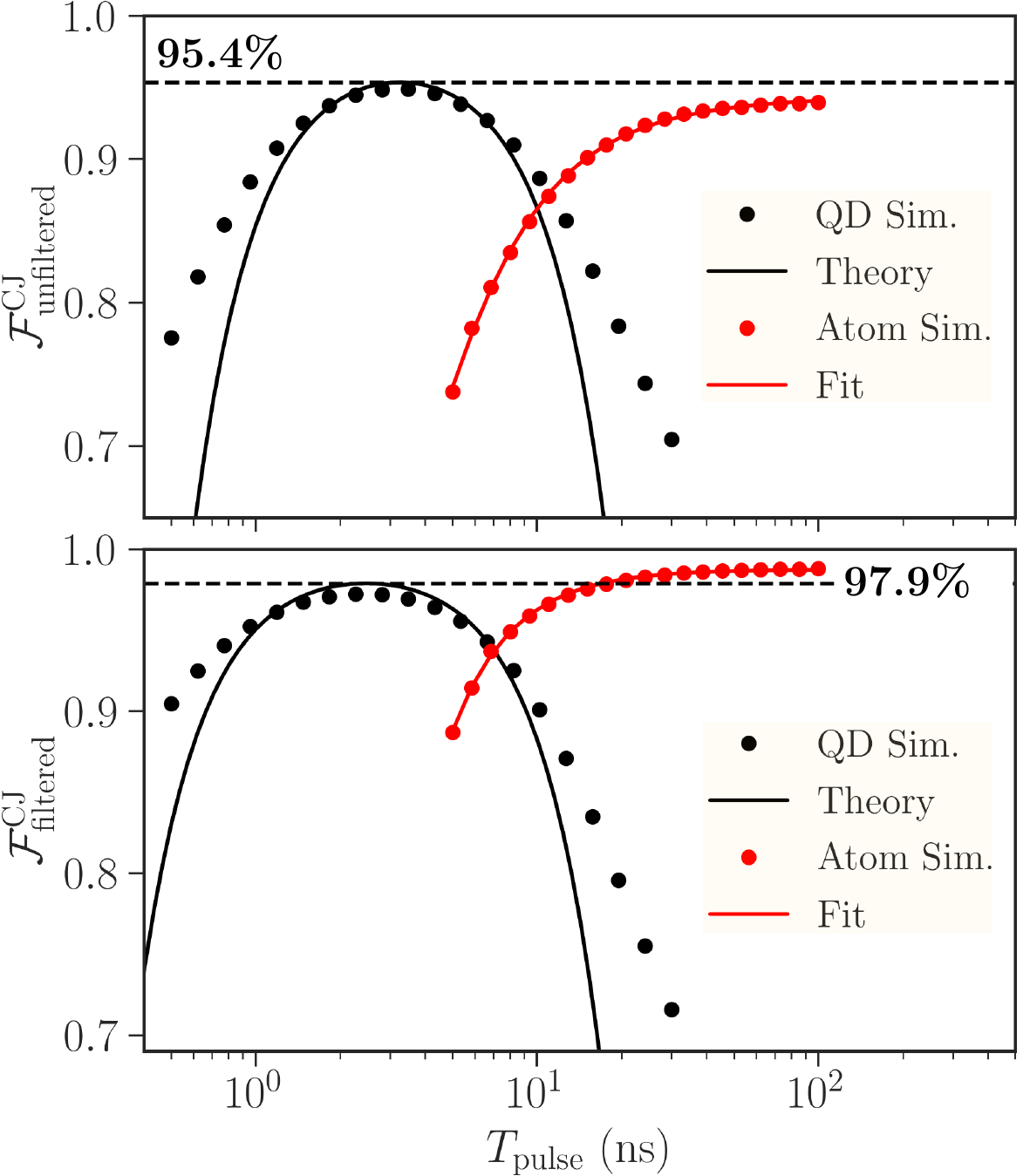}
	\caption{(color online) Comparison of average fidelities between QD (black) and Rb-atom systems (red) as a function of the input pulse duration $T_{\text{pulse}}$. 
	Symbols denote numerical results. The full lines show the analytical results for the QDs and a simple fit for the Rb data.
	Dashed lines indicate the maximum achievable fidelity for the QD system predicted by Eq.~(\ref{eq:3def}). The parameters of the simulation are provided in the main text.}
	\label{fig:comparison}
\end{figure}

Finally, in order to reduce the probability of off-resonant spin flip $\Gamma^2/\Delta^2$, a strong external magnetic field could be applied to increase the ground-state splitting. In Ref.~\cite{Appel2020}, a low in-plane $g$-factor of 0.26 for the XP exciton was observed, leading to a sufficiently large splitting of $\Delta/2\pi = 7.3$~GHz at 2T while preserving good spin coherence properties. This ensures $\Delta/\Gamma\approx20$ with $\mathcal{F}^{\text{CJ}}_{\text{unfiltered}}$ approaching $99.9\%$. 

To gauge the full performance of the protocol we now combine all imperfections of the protocol with realistic parameters in the QD-waveguide system: $\sigma_e = 0.1$~ns$^{-1}$, $\Gamma = 2.48$~ns$^{-1}$, $\gamma = 0.05$~ns$^{-1}$, $\gamma_d = 0.03$~ns$^{-1}$, $\Delta/2\pi = 7.3$~GHz, $T_2^* = 21.4$ ns~\cite{Appel2020} and assume $\varepsilon=0$. With these parameters  $\mathcal{F}^{\text{CJ}}_{\text{unfiltered}}$ ($\mathcal{F}^{\text{CJ}}_{\text{filtered}}$) reaches 95.4\% (97.9\%) with a success probability of 98\% (95.1\%) (Fig.~\ref{fig:comparison}).

Finally, we compare the performance of a waveguide-embedded QD to that of a $^{87}$Rb-atom trapped next to a fiber-coupled ultrahigh-Q microtoroid whispering-gallery-mode (WGM) resonator. A transverse magnetic (TM) mode of the resonator is tuned to be resonant with the $F=1\rightarrow F’=1$ transition of the D1 line of $^{87}$Rb. The $\Lambda$-system is comprised by the two ground states $F=1,m_F=\pm1$, defining the atomic qubit, and the excited state $F’=1,m_F=0$. The evanescent field of the clockwise (counterclockwise) TM mode, associated with left-(right-) propagating light in the fiber, is coupled primarily to the $\sigma^{+}$ ($\sigma^{-}$) transition of the atom~\cite{junge2013strong,aiello2015transverse,Lodahl2017}, allowing to independently address each transition in the $\Lambda$-system. By applying a weak external magnetic field of $20G$, we induce a Zeeman energy shift to the ground states that lifts the degeneracy in frequency between the two transitions, making the photonic qubit both polarization- and frequency-encoded. We simulate numerically the photon-to-atom state-transfer fidelity using the following realistic parameters~\cite{shomroni2014,rosenblum2016,Bechler2018} : coherent atom-cavity coupling rate of $g=2\pi\times16\text{ MHz}$, fiber-cavity coupling rate of $\kappa_{ex}=2\pi\times30\text{ MHz}$, intrinsic cavity loss rate of $\kappa_{i}=2\pi\times2\text{ MHz}$ (corresponding to $Q\approx 10^8$), atomic free-space amplitude decay rate of $\gamma=2\pi\times3\text{ MHz}$, parasitic coupling rate between the two modes of the cavity of $h=2\pi\times1\text{ MHz}$, and an undesired polarization component $r_{\sigma} = 0.19$ (see Ref.~\cite{Rosenblum2017}). The simulation also takes into account atomic transitions outside of the $\Lambda$-system that affect the ideal operation of the scheme. Dephasing processes in trapped single atoms are reported to be on the order of $100\text{ }\mu$s~\cite{reitz2013coherence,Reiserer2015} and thus have a negligible impact on the fidelity when $T_{\text{pulse}}<100$~ns.

When comparing between the two platforms, one must keep in mind that the waveguide-QD system is stationary whereas trapping and cooling of a single atom next to a WGM resonator remains a challenging task~\cite{will2021coupling,zhou2021subwavelength}. Simulations show that both systems have a similar maximum fidelity, yet their respective optimal pulse durations differ by about two orders of magnitude (Fig.~\ref{fig:comparison}). The fidelity in the atom-cavity system approaches its maximum at a pulse duration of around $100\text{ ns}$, as opposed to an optimal pulse duration of $3.49\text{ ns}$ for the QD-waveguide platform, which is dictated by the ratio between the emitter's spin dephasing time and its decay rate, $T_{\text{optimal}}\propto\sqrt{T^*_2/\Gamma}$. A shorter pulse duration ($T_{\text{pulse}}\leq10 \text{ ns} $) is advantageous for increasing the repetition rate of protocols, e.g. boosting the rate of memory-assisted measurement-device-independent quantum key distribution~\cite{Panayi2014,Boone2015}. In this regime, the QD-waveguide platform is favorable but care should be taken to preserve the spin coherence. On the other hand, a longer pulse duration ($T_{\text{pulse}}>10 \text{ ns} $), suitable to the atom-cavity platform, is favorable for interfacing with low-bandwidth emitters. Additionally, the atomic system is advantageous for applications requiring long storage times thanks to its significantly lower dephasing rates.


\section{Conclusion}
We have proposed a passive scheme to perform deterministic quantum state transfer from a frequency-encoded photon to a quantum-dot spin mediated by a nanophotonic waveguide. Strikingly, with the exception of pure dephasing, we find that the state-transfer fidelity is insensitive to first order in the small parameters for the considered spectral, coupling  and spin dephasing errors. This demonstrates the robustness of the scheme. 
The thorough fidelity analysis unravels the influence of various physical processes governing the quality of the quantum state transfer and  hence provides an important guideline for experimental realizations with QDs as well as other emitters.

The photon-spin transfer protocol is vital for applications in quantum-information processing with frequency-encoded qubits. In particular, the coherent exchange of arbitrary states between the photon and the emitter enables deterministic SWAP gates \cite{Bechler2018}, quantum non-demolition detection and memory-assisted satellite quantum key distribution~\cite{Boone2015,Anton2020}. These functionalities are crucial for implementing scalable quantum networks.

\section*{Acknowledgements}
M.L.C, A.T., P.L., and A.S.S. acknowledge financial support from Danmarks Grundforskningsfond (DNRF 139, Hy-Q Center for Hybrid Quantum Networks) and the European Union’s Horizon 2020 Research and innovation Programme under the Marie Sklodowska-Curie Grant Agreements No.~861097 (QUDOT-TECH) and No.~820445 (Quantum Internet Alliance). B.D acknowledges support from the Israeli Science Foundation, the Binational Science Foundation, H2020 Excellent Science (DAALI, 899275), the Minerva Foundation, and a
research grant from Dr. Saul Unter.

\newpage
\onecolumngrid
\section*{Appendix A: Formula of the average gate fidelity in heralded operation}

\label{sec:linearr}
In this section we derive the following linear relation between the Choi-Jamiolkowski fidelity $\mathcal{F}^{\text{CJ}}$ and the weighted average conditional fidelity $\mathcal{\bar{F}}^{(c)}$:
\begin{align}
    \mathcal{\bar{F}}^{(c)} = \frac{2}{3}\mathcal{F}^{\text{CJ}}+\frac{1}{3}.
\end{align}
The proof is made by deriving an expression for each fidelity and comparing their results. 
\newline\\
\textit{\underline{Proof:}} 
\newline
\paragraph*{\bf{(1) Choi-Jamiolkowski fidelity $\mathcal{F}^{\text{CJ}}$.}} The starting point for the Choi-Jamiolkowski fidelity is to consider a (fictitious) input state, which is a Bell-state between two sub-systems A and S:
\begin{align}
    \ket{\psi_{\text{in}}}=\frac{1}{\sqrt{2}}\bigg(\ket{0_A 0_S}+\ket{1_A 1_S}\bigg).
\end{align}
The Choi-Jamilkowski fidelity corresponds to the fidelity of the state when we apply our map $\mathcal{I}_A \otimes \mathcal{E}_S$ to the  ideal EPR-pair in the bipartite system. If the Choi-Jamilkowski fidelity is above 50\% this guarantees that the fictitious state would remains entangled after the (conditional) operation on qubit S, and thus signifies that the map is entanglement-preserving. 

The quantum gate process in general is modelled by the superoperator $\mathcal{E}$ acting on an outer product $\ket{i}\bra{j}$.
Requiring the density matrix to be normalized after a non-trace preserving map, i.e. by conditioning on a photon in the output,  the density matrix of the output state is
\begin{align}
    \rho_{\text{out}}&=\frac{\ket{\psi_{\text{out}}}\bra{\psi_{\text{out}}}}{\Tr(\ket{\psi_{\text{out}}}\bra{\psi_{\text{out}}})}\nonumber\\
    &=\frac{\ket{\psi_{\text{out}}}\bra{\psi_{\text{out}}}}{\sum\limits_{i,j\in\{0,1\}}^{}\bra{i_A j_S}(\ket{\psi_{\text{out}}}\bra{\psi_{\text{out}}})\ket{i_A j_S}}\nonumber\\
    &=\frac{\ket{0_A}\mathcal{E}(\ket{0_S}\bra{0_S})\bra{0_A}+\ket{0_A}\mathcal{E}(\ket{0_S}\bra{1_S})\bra{1_A}+\ket{1_A}\mathcal{E}(\ket{1_S}\bra{0_S})\bra{0_A}+\ket{1_A}\mathcal{E}(\ket{1_S}\bra{1_S})\bra{1_A} }{\sum\limits_{j,m=\{0,1\}}^{}\mathcal{M}_{j,m,m,j} },
\end{align}
where $\mathcal{M}_{k,l,l,k}=\bra{k_S}\mathcal{E}(\ket{l_S}\bra{l_S})\ket{k_S}$ can be interpreted as the probability for subsystem S to be in state $\ket{k}$  after the operation $\mathcal{E}$ when starting in state $\ket{l}$.

For a perfect gate operation denoted by the unitary operator $\mathcal{U}_{\text{ideal}}$, we have
\begin{align}
    \frac{1}{\sqrt{2}}\bigg(\ket{0_A 0_S}+\ket{1_A 1_S}\bigg) \quad&\to\quad 
    \frac{1}{\sqrt{2}}\bigg[ \ket{0_A}\otimes\mathcal{U}_{\text{ideal}}\ket{0_S}+\ket{1_A}\otimes\mathcal{U}_{\text{ideal}}\ket{1_S} \bigg]=\ket{\psi_{\text{ideal}}}.
\end{align}
Therefore, the Choi-Jamiolkowski fidelity of the mapping  is
\begin{align}
    \boxed{\mathcal{F}^{\text{CJ}}=\bra{\psi_{\text{ideal}}}\rho_{\text{out}}\ket{\psi_{\text{ideal}}}
    = \frac{1}{2}\frac{\sum\limits_{j,m=\{0,1\}}^{}\mathcal{M'}_{j,j,m,m} }{\sum\limits_{j,m=\{0,1\}}^{}\mathcal{M}_{j,m,m,j}},}\label{eq:cjjj}
\end{align}
where $\mathcal{M'}_{k,l,i,j}=\bra{k_S}\mathcal{U}^\dagger_{\text{ideal}}\mathcal{E}(\ket{l_S}\bra{i_S})\mathcal{U}_{\text{ideal}}\ket{j_S}$ computes the overlap with the ideal state. 

 \paragraph*{\bf{(2) Weighted average of the conditional fidelity $\mathcal{\bar{F}}^{(c)}$.}} To evaluate the weighted average we need to consider the evolution of specific states. 

The input-output relation for applying a unitary operator $\mathcal{U}$ on an arbitrary input qubit S on the Bloch sphere is
\begin{align}
    \ket{\psi_{\text{in}}}=\cos{\frac{\theta}{2}}\ket{0_S} + \sin{\frac{\theta}{2}}e^{i\phi}\ket{1_S} \quad \xrightarrow{\mathcal{U}} \quad \cos{\frac{\theta}{2}}\mathcal{U}\ket{0_S} + \sin{\frac{\theta}{2}}e^{i\phi}\mathcal{U}\ket{1_S}=\ket{\psi_{\text{out}}},
\end{align}
which can be generalized to a non-unitary operation $\mathcal{E}$ via $\mathcal{U}\ket{i}\bra{j}\mathcal{U}^\dagger\to\mathcal{E}(\ket{i}\bra{j})$. The normalized density matrix of the output state then becomes
\begin{align}
    \rho_{\text{out}}&=\frac{\ket{\psi_{\text{out}}}\bra{\psi_{\text{out}}}}{\Tr(\ket{\psi_{\text{out}}}\bra{\psi_{\text{out}}})}\nonumber\\
    &= \frac{\cos^2(\frac{\theta}{2})\mathcal{E}(\ket{0_S}\bra{0_S}) + \sin^2(\frac{\theta}{2})\mathcal{E}(\ket{1_S}\bra{1_S}) + \sin(\frac{\theta}{2})\cos(\frac{\theta}{2})e^{i\phi}\mathcal{E}(\ket{1_S}\bra{0_S}) + \text{H.c.}}{\sum\limits_{i=\{0,1\}}^{}\bra{i_S}\bigg(\cos^2(\frac{\theta}{2})\mathcal{E}(\ket{0_S}\bra{0_S}) + \sin^2(\frac{\theta}{2})\mathcal{E}(\ket{1_S}\bra{1_S}) + \sin(\frac{\theta}{2})\cos(\frac{\theta}{2})e^{i\phi}\mathcal{E}(\ket{1_S}\bra{0_S})+\text{H.c.}\bigg)\ket{i_S}},
\end{align}
where the denominator is the success probability $P^s_i$ of the map for  the specific input qubit state. The fidelity for each cardinal input state can then be computed using $\mathcal{F}^{(c)}_i=\bra{\psi_{\text{ideal}}}\rho_{\text{out}}\ket{\psi_{\text{ideal}}}$ at 6 different sets of values for $\theta$ and $\phi$.
When evaluating the weighted average $\mathcal{\bar{F}}^{(c)}$, the weighted sum $\sum_i P^s_i \mathcal{F}^{(c)}_i$ and $\sum_i P^s_i$ will contain the sum of phases $e^{\pm i\phi}$ over $4$ different sets of azimuthal angle $\phi$, thus the phases will eventually be cancelled out according to the table below:
\begin{table*}[ht]
    \centering
 \begin{tabular}{|c|c| c| c| c|c| c| c| c|}
 \hline
Index $i$ & $\theta$ &$\phi$ & $e^{i\phi}$& $e^{2i\phi}$& $e^{-i\phi}$& $e^{-2i\phi}$&  Fidelity\\ 
 [0.6ex] 
 \hline\hline
 1&$0$ & - & - & - & - & - &$\mathcal{F}^{(c)}_1$ \\
\hline
 2&$\pi$ & - & - & - & - & - &$\mathcal{F}^{(c)}_2$ \\
 \hline
  3&$\pi/2$ & 0 & 1 & 1 & 1 & 1  &$\mathcal{F}^{(c)}_3$\\
  \hline
  4&$\pi/2$ & $\pi$ & $-1$ & 1 & $-1$ & 1  &$\mathcal{F}^{(c)}_4$\\
  \hline
  5&$\pi/2$ & $\pi/2$ & $i$ & $-1$ & $-i$ & $-1$  &$\mathcal{F}^{(c)}_5$\\
  \hline
  6&$\pi/2$ & $3\pi/2$ & $-i$ & $-1$ & $i$ & $-1$  &$\mathcal{F}^{(c)}_6$\\
  \hline
 \end{tabular}
\end{table*}
\newline
The weighted average conditional fidelity is therefore
\begin{align}
    \boxed{\mathcal{\bar{F}}^{(c)} = \frac{\sum_i P^s_i \mathcal{F}^{(c)}_i }{\sum_i P^s_i} = \frac{1}{3}\frac{\sum\limits_{j,m=\{0,1\}}^{}\bigg(\mathcal{M'}_{j,j,m,m}+\mathcal{M'}_{j,m,m,j}\bigg) }{\sum\limits_{j,m=\{0,1\}}^{}\mathcal{M}_{j,m,m,j}}  .}
    \label{eq:wac}
\end{align}

\paragraph*{\bf{(3) Comparing two fidelities.}}
Comparing~(\ref{eq:cjjj}) with (\ref{eq:wac}) we find a relation between the two expressions
\begin{align}
    \mathcal{\bar{F}}^{(c)} = \frac{2}{3}\mathcal{F}^{\text{CJ}} + \frac{1}{3}\frac{\sum\limits_{j,m=\{0,1\}}^{}\mathcal{M'}_{j,m,m,j}}{\sum\limits_{j,m=\{0,1\}}^{}\mathcal{M}_{j,m,m,j}},
    \label{eq:linear1}
\end{align}
Using the property that the trace of a matrix is invariant under the unitary transformation $\mathcal{U}_{\text{ideal}}$, we arrive at
\begin{align}
    \boxed{\mathcal{\bar{F}}^{(c)} = \frac{2}{3}\mathcal{F}^{\text{CJ}} + \frac{1}{3}.}
    \label{eq:linear2}
\end{align}
Eq.~(\ref{eq:linear2}) thus establishes a linear relation between two fidelity definitions, which allows us to extract the Choi-Jamiolkowski fidelity for the heralded state-transfer protocol.

\twocolumngrid
\bibliography{reflist}

\end{document}